\documentclass[lettersize,journal]{IEEEtran}
\usepackage{amsmath,amsfonts}
\usepackage{algorithmic}
\usepackage{array}
\usepackage[caption=false,font=normalsize,labelfont=sf,textfont=sf]{subfig}
\usepackage{textcomp}
\usepackage{stfloats}
\PassOptionsToPackage{hyphens}{url}\usepackage{hyperref}
\usepackage{url}
\usepackage{verbatim}
\usepackage{amsmath, mathtools, nccmath,amsthm,amsfonts,amssymb,hyperref,url}
\usepackage{float}

\usepackage{longtable}
\usepackage{graphicx,subfig}
\usepackage{xcolor}
\usepackage{stfloats}
\usepackage{rotating} 
\usepackage{tikz}
\usepackage{blindtext}
\DeclareMathOperator*{\argmax}{{arg\;max}}

\usepackage{soul,latexsym,amsmath,float,caption,amsfonts,amssymb,txfonts,enumitem,mathtools,amsthm,enumitem,physics}
\usepackage{color}
\usepackage{graphicx}
\usepackage{cite}

\usepackage{array}
\usepackage{multicol}
\usepackage[bottom]{footmisc}
\usepackage{url}
\usepackage{wrapfig}
\usepackage{capt-of}
\usepackage{caption}
\usepackage{etoolbox}
\usepackage[ruled,vlined,linesnumbered]{algorithm2e}
\usepackage{comment}

\usetikzlibrary{shapes.geometric, arrows}
\usepackage{multirow}

\usepackage{chngcntr}
\usepackage{framed}
\usepackage{graphicx}
\usepackage{balance}
\usepackage[normalem]{ulem}
\usepackage{makecell}

\begin{document}

\title{Multi-Agent Reinforcement Learning for UAV-Based Chemical Plume Source Localization}

\author{Zhirun Li,~\IEEEmembership{Student Member,~IEEE}, Derek Hollenbeck,~\IEEEmembership{Member,~IEEE}, Ruikun Wu,~\IEEEmembership{Student Member,~IEEE}, Michelle Sherman,~\IEEEmembership{Member,~IEEE}, Sihua Shao,~\IEEEmembership{Senior Member,~IEEE}, Xiang Sun,~\IEEEmembership{Member,~IEEE},\\ and Mostafa Hassanalian,~\IEEEmembership{Member,~IEEE} \vspace{-20pt}

\thanks{\textit{Corresponding author: Sihua Shao.}}

\thanks{Z. Li is with the Department of Electrical and Computer Engineering, University of New Mexico, Albuquerque, NM 87131 USA (e-mail: zhirunli@unm.edu).}

\thanks{D. Hollenbeck is with the Department of Mechanical Engineering, University of California, Merced, CA 95343 USA (e-mail: dhollenbeck@ucmerced.edu).}

\thanks{R. Wu is with the Department of Electrical Engineering, Colorado School of Mines, Golden, CO 80401 USA (e-mail: ruikun\_wu@mines.edu).}

\thanks{M. Sherman is with the Department of Electrical Engineering, New Mexico Tech, Socorro, NM 87801 USA (e-mail: michelle.sherman@student.nmt.edu).}

\thanks{S. Shao is with the Department of Electrical Engineering, Colorado School of Mines, Golden, CO 80401 USA (e-mail: sihua.shao@mines.edu).}

\thanks{X. Sun is with Autonomous Solutions, Inc., Mendon, UT 84325 USA (e-mail: xiangsun@ieee.org).}

\thanks{M. Hassanalian is with the Department of Mechanical Engineering, New Mexico Tech, Socorro, NM 87801 USA (e-mail: mostafa.hassanalian@nmt.edu).}
}

\maketitle

\begin{abstract}
Undocumented orphaned wells pose significant health and environmental risks to nearby communities by releasing toxic gases and contaminating water sources, with methane emissions being a primary concern. Traditional survey methods such as magnetometry often fail to detect older wells effectively. In contrast, aerial in-situ sensing using unmanned aerial vehicles (UAVs) offers a promising alternative for methane emission detection and source localization. This study presents a robust and efficient framework based on a multi-agent deep reinforcement learning (MARL) algorithm for the chemical plume source localization (CPSL) problem. The proposed approach leverages virtual anchor nodes to coordinate UAV navigation, enabling collaborative sensing of gas concentrations and wind velocities through onboard and shared measurements. Source identification is achieved by analyzing the historical trajectory of anchor node placements within the plume. Comparative evaluations against the fluxotaxis method demonstrate that the MARL framework achieves superior performance in both localization accuracy and operational efficiency.
\end{abstract}

\begin{IEEEkeywords}
chemical plume source localization (CPSL), formation control, cooperative sensing, multi-agent deep reinforcement learning
\end{IEEEkeywords}

\section{Introduction}
\IEEEPARstart{U}{ndocumented} orphaned wells can remain undetected and uncharacterized for years or even decades, potentially leaking methane and other potent greenhouse gas into the atmosphere. Estimates suggest that over 126,000 orphaned wells exist, along with an additional 310,000 to 800,000 undocumented orphaned wells across 32 oil- and gas-producing states \cite{IOGCC2021}. Although technologies are available to aid in locating, identifying and characterizing these wells, traditional aerial survey methods, such as magnetometry, often prove ineffective---particularly for older wells or those in poor condition \cite{DOE_Catalog}. In some cases, undocumented orphaned wells have had their casings removed for repurposing, leaving them backfilled or ``plugged" with readily available materials. To better mitigate atmospheric emissions and other environmental hazards, more robust well identification and characterization techniques are needed. The Department of Energy (DOE) Consortium Advancing Technology for Assessment of Lost Oil \& Gas Wells (CATALOG) is actively exploring non-invasive methods to address these challenges \cite{DOE_Catalog}.

\begin{figure}[t]
   \centering
    \includegraphics[width=0.45\textwidth]{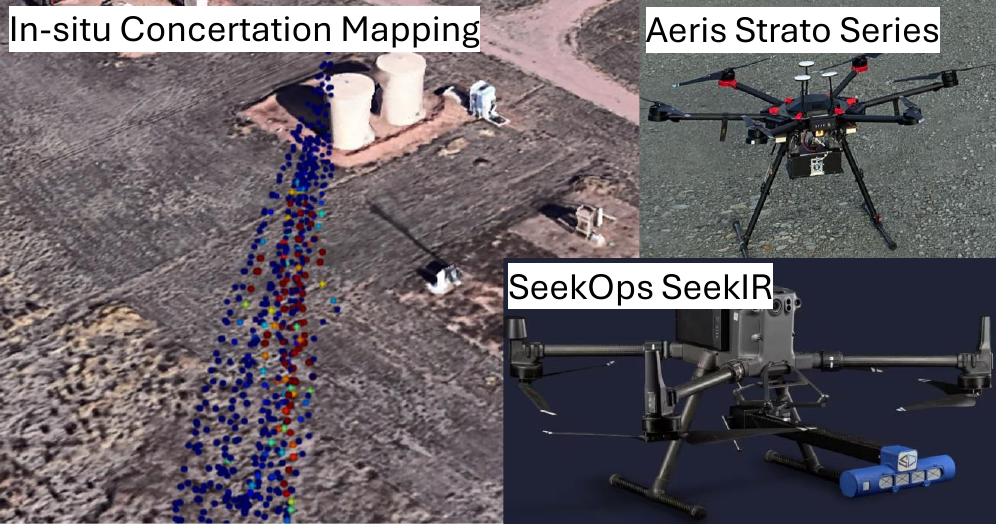}
    \caption{In-situ gas concentration measurement results and two UAV-sensor products \cite{aeris_strato,seekops_technology}.}
    \label{fig:UAV_and_sensor}
\end{figure}

Methane emissions from undocumented orphaned wells are generally low in volume. A study of 568 orphaned wells across U.S. states (OH, WY, UT, CO, PA, and WV) and Canadian provinces (NB and BC) found methane emission rates ranging from 1.8 × 10$^{-3}$ to 48 grams per hour per well, varying based on plugging status, well type, and location, with an overall average of 6 grams/hour \cite{williams2020methane}. Notably, wells in the top 10\% of emission rates ($>$10 grams per hour) accounted for 91\% of total emissions.

Current remote sensing technologies, such as MathaneSAT \cite{MethaneSAT} and fixed-wing aircraft surveys \cite{souza2023framework}, lack the sensitivity needed to detect these low emission rates. Ground-based direct measurement techniques, including flux chambers and high-flow samplers \cite{VentBusters}, require prior knowledge of well locations, accessible wellheads, and extensive deployment efforts. In contrast, an in-situ unmanned aerial vehicle (UAV) sensing solution (see Fig.~\ref{fig:UAV_and_sensor}) employs open- or closed-path laser absorption spectroscopy \cite{Axetris_LGD,aeris_strato,seekops_technology} to detect gas concentrations at the part-per-billion (ppb) level. This approach enables UAVs to access difficult-to-reach regions and detect gas concentrations over up to 50 meters downwind from a source emitting as little as 10 grams per hour. Equipped with highly sensitive in-situ sensors, multiple UAVs can collaboratively share sensor data to conduct efficient aerial scans, detect chemical signatures, and trace those signatures upwind to pinpoint the emitter’s location---a process known as chemical plume source localization (CPSL). Once the emitter's location is identified, the same UAV team can immediately begin sampling at various points downwind of the emitter. By applying suitable dispersion models, they can accurately quantify the emission rate \cite{hollenbeck2022single}.

\begin{figure}[t]
   \centering
    \includegraphics[width=0.48\textwidth]{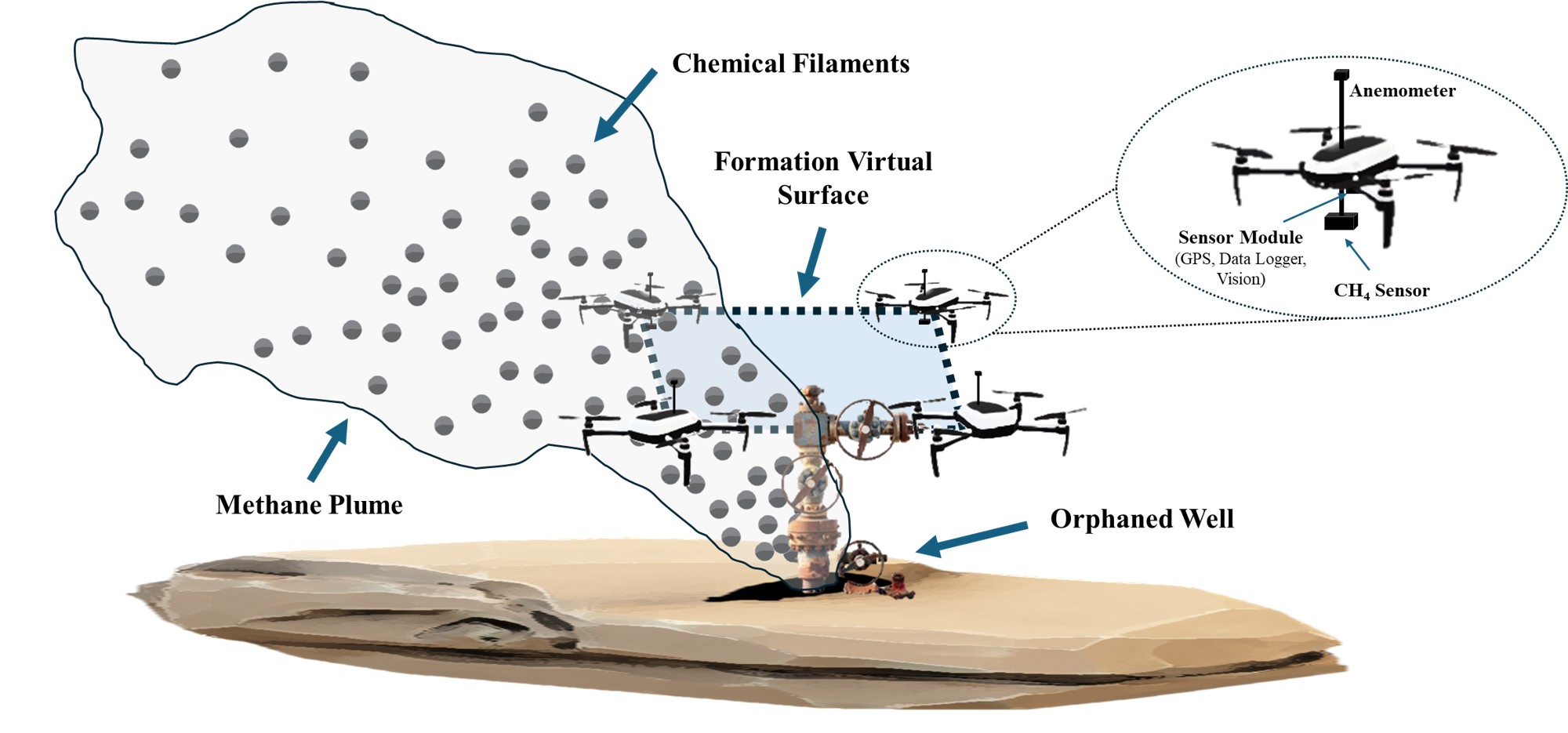}
    \caption{Envisioned Multi-UAV CPSL scenario.}
    \label{fig:IntroFig}
\end{figure}

This work focuses on the CPSL mission, which aims to identify the origin of a chemical leak by tracing its dispersing plume. In aerial plume tracing, UAVs equipped with chemical sensors navigate the air to detect and follow chemical concentration cues, often in conjunction with wind measurements, to locate the source. \textit{This task is challenging due to the turbulent and patchy nature of atmospheric plumes in high Reynolds number flows, which results in an intermittent chemical signal rather than a smooth gradient \cite{vergassola2007infotaxis}.} To address issues such as noisy, sporadic measurements, uncertain wind conditions, and large search areas, we propose a centralized training and decentralized execution (CTDE) framework using a multi-agent deep reinforcement learning (MARL) algorithm. Our approach efficiently localizes an emitter while accounting for practical considerations, including sensor noise, wind turbulence, UAV dynamics, collision avoidance, formation control, and source declaration conditions, making it suitable for real-time implementation. The envisioned multi-UAV scenario is illustrated in Fig.~\ref{fig:IntroFig}. The agents execute the three key CPSL phases: seek, trace, and declare. In the source declaration phase, the emitter's location is estimated based on the centroid of the virtual surface formed by the UAVs.

\textit{Contributions:} This work makes four key contributions: 
\begin{itemize}
\item We identify the key challenges in developing a cooperative sensing strategy for CPSL, focusing on environmental stochasticity, formation control, and collision avoidance.

\item We develop an efficient and robust MARL framework for real-world complexities, including wind dynamics and transient sensor readings.

\item The proposed MARL framework is evaluated against the fluxotaxis method and demonstrates superior performance in both efficiency and accuracy.

\item We use animation-based visualization to assess plume detection, tracing, and localization behaviors of the multi-UAV formation. 

\end{itemize}

\textit{Organization:} Sec.~\ref{sec:Ch5-Related} provides a comprehensive review of CPSL methods in aerial plume tracing. Sec.~\ref{sec:Ch5-Tasks} outlines the three phases of the CPSL task. Sec.~\ref{sec:ch5-System Model} details the system model. Sec.~\ref{sec:Ch5-MADRL} describes the proposed MARL framework. Simulation results are analyzed in Sec.~\ref{sec:Ch5-NumericalResults}, and Sec.~\ref{sec:Ch5-Conclusion} concludes the article with planned future work.

\section{Related Work} \label{sec:Ch5-Related}
Current CPSL strategies can be broadly classified into four main approaches: reactive bio-inspired methods, probabilistic and model-based methods, multi-robot and swarm methods, and learning-based methods.

\textit{Reactive bio-inspired methods} mimic odor-tracking behaviors observed in biological organisms such as insects and bacteria. These strategies rely on local sensor stimuli to guide movement. For instance, \textit{chemotaxis} directs the agent up a concentration gradient, similar to bacteria navigating toward higher nutrient levels \cite{staples2023comparison,spears2012physicomimetics}. Likewise, flying insects employ \textit{anemotaxis}, moving upwind upon detecting odor, as the source is typically located upwind in a dispersing plume  \cite{staples2023comparison,spears2012physicomimetics}. Bio-inspired reactive algorithms are simple, fast, and computationally lightweight, making them well-suited for resource-constrained UAVs \cite{shigaki2023robust}. They perform effectively in laminar or moderate flow conditions where plumes remain coherent \cite{zarzhitsky2004agent}. However, these methods struggle in highly turbulent, patchy plumes typical of real-world environments. A purely chemotactic robot may become trapped in local concentration peaks and fail near the source due to misleading fluctuations caused by turbulent eddies. Similarly, an upwind surge strategy may mislead the robot into a false chase if it detects a transient odor signal from a swirling eddy, only to lose it due to shifting wind. As the delay between detections increases, the performance of reactive algorithms declines \cite{staples2023comparison}. Lacking memory of past detections, these robots may wander inefficiently or repeatedly overshoot the source in intermittent plumes.

\textit{Probabilistic and model-based methods} maintain an internal representation of the source location, continuously updating it based on sensor readings \cite{staples2023comparison}. Unlike purely reactive strategies, these methods incorporate both current and past information to plan the robot's path. Common approaches include \textit{occupancy grid mapping} \cite{jakuba2007stochastic}, which estimates the probability of a source in different spatial regions, \textit{particle filters} \cite{neumann2013gas}, and \textit{Bayesian inference} \cite{1703649} for source estimation. Information-theoretic search methods, such as \textit{Infotaxis}, treat each detection (or non-detection) as information, guiding movement to maximize expected information gain about the source \cite{vergassola2007infotaxis}. These techniques often rely on dispersion models (e.g., Gaussian plume models) or statistical assumptions to update a likelihood map of the source location. Probabilistic approaches are systematic and highly robust to uncertainty, making them effective in environments where detections are sparse or intermittent---scenarios where reactive methods typically fail. By retaining memory of previous detections, these methods can infer the likely source location even from limited observations. However, their computational demands pose challenges, especially for real-time deployment on small UAVs. Calculating and updating probability maps or solving partially observable decision problems can be computationally intensive. While information-theoretic control is theoretically optimal, it becomes intractable in large state spaces with limited onboard processing power \cite{dawson2021hybrid}.

\textit{Multi-robot and swarm methods} leverage multiple UAVs for plume tracing, significantly accelerating source localization. These strategies coordinate a team of agents that explore and exchange information concurrently. Some approaches use swarm intelligence, where simple local rules lead to cooperative behavior. For instance, \textit{particle swarm optimization} (PSO) models UAVs as particles attracted to higher concentrations while repelling each other to maximize coverage \cite{wang2022odor,CHEN2019123}. A notable example is the \textit{physicomimetics} approach, which arranges UAVs in a lattice formation to act as distributed sensors, computing chemical concentration gradients through the \textit{fluxotaxis} algorithm \cite{spears2012physicomimetics, hollenbeck2022digital}. In practical applications, multiple UAVs can divide the search space, follow plume filaments from different entry points, or execute systematic grid searches in parallel. The primary advantage of multi-agent plume tracing is its speed and efficiency through parallelism. Additionally, multi-robot systems enhance resilience and robustness: even if one UAV loses the plume or malfunctions, others can continue the search, ensuring mission success \cite{hinsen2023exploration}. However, coordinating multiple agents presents several challenges, including communication overhead, collision avoidance, and downwash interference to nearby UAVs. Cost is also a significant factor---one Aeries Strato Series UAV-Sensor system (Fig.~\ref{fig:UAV_and_sensor}) exceeds \$60K. From an algorithmic standpoint, multi-agent search dramatically increases the state space, making optimal planning computationally prohibitive (``curse of dimensionality'' for centralized solutions) \cite{dawson2021hybrid}. Decentralized heuristics help mitigate this issue but require careful design to prevent suboptimal behaviors, such as all UAVs converging on the same false lead.

\textit{Learning-based methods} have emerged as powerful approaches for CPSL with advancements in AI. In particular, reinforcement learning (RL) enables robots to learn plume-tracing policies through trial and error in simulations or real-world experiments. Instead of manually coding surge-and-cast behaviors or defining an information gain function, the robot’s controller---often a deep neural network in deep RL \cite{sherman2024counter}---is trained to process sensor inputs (e.g., recent concentration readings, wind direction) and generate movement actions that maximize a reward \cite{wang2023learn,10466498,ZHAO202267}. MARL extends this framework by training multiple agents to collaborate toward a shared goal. These learning methods can be model-free, meaning they do not rely on an explicit plume model. The key advantage of learning-based approaches is their adaptability. An RL agent can, in theory, learn optimal behaviors for given conditions by training on a diverse set of plume scenarios. Additionally, learned policies can exhibit robustness to sensor noise and delays, as the agent incorporates these uncertainties into its decision-making process if they were present during training. Once trained, executing the learned policy is computationally lightweight---requiring only a feed-forward pass through a neural network---making it feasible for deployment on small UAVs. However, deep RL-based methods require extensive training data and careful hyperparameter tuning, which can be a significant design challenge. To address this, we utilize a filament-based dispersion model \cite{farrell2002filament} to simulate realistic, time-varying plume dynamics. Additionally, we are developing a digital twin model \cite{hollenbeck2022single} to match simulated outputs with controlled release experiments, improving the fidelity of the training environment. To mitigate the sample inefficiency of RL, we incorporate an upwind bias as a prior and hybridize RL with information-driven rewards. Furthermore, we employ the CTDE technique to enhance training stability in multi-agent systems and use visualization of different learning phases to interpret behavioral performance.

Compared to classical bio-inspired or probabilistic algorithms, MARL provides several key advantages for aerial plume tracing. These include implicit coordination and adaptation among agents, improved robustness to turbulence and uncertainty, global optimization of search objectives, and a reduced reliance on manual tuning and explicit plume models.

\section{Phases of the Chemical Plume Source Localization Problem via Multi-UAVs}
\label{sec:Ch5-Tasks}
In this work, we focus on a single-emitter scenario, which is a representative case for undocumented orphaned wells. The UAVs begin their mission with no prior knowledge of the emitter’s location and proceed through three sequential phases---seek, track, and localize---terminating upon either satisfying a defined stopping condition or reaching a preset time limit. Light to moderate wind turbulence is assumed, which is a practical consideration given that UAV deployment can be scheduled based on favorable weather conditions. All UAVs are assumed to remain within communication range of a ground control station, enabling real-time exchange of sensor readings and location data. Solving the CPSL problem requires balancing robustness with efficiency. Our approach focuses on tracking transient chemical signatures using real-time onboard sensors, guiding UAVs to progressively detect, trace, and ultimately localize the emitter. The CPSL task is structured into the following three phases:



\subsubsection{Seek Phase} \textit{Search} for initial chemical signatures.

Initially, UAVs are assumed to start outside the chemical plume. Their primary objective in this phase is to explore the environment in search of detectable chemical concentrations above a predefined threshold to distinguish it from background noise. This phase resembles the classic casting behavior described in plume-tracking literature \cite{spears2012physicomimetics}, where agents scan the environment systematically. Traditional multi-UAV casting strategies can be energy-inefficient due to overlapping paths and excessive area coverage. In our approach, UAVs are uniformly distributed along the horizontal ($X$) axis and perform vertical sweeps across the search grid. This strategy leverages the wide horizontal cross-section of the plume, making detection more efficient and scalable.

\subsubsection{Trace Phase} \textit{Trace} the chemical plume. 

Upon initial detection, UAVs transition into the trace phase, sharing local sensor data such as methane concentration ($\rho$) and wind velocity ($\mathbf{V}$) to collectively track the plume. The goal is to guide neighboring UAVs into the plume region, facilitating coordinated upwind movement. While traditional plume tracing strategies---such as chemotaxis, anemotaxis, and fluxotaxis---rely on instantaneous measurements to determine movement \cite{spears2012physicomimetics,hollenbeck2023swarm}, our method improves robustness by using a running average over the past $\tau_m$ samples of sensor data. This smoothing technique reduces the impact of transient fluctuations and allows UAVs to respond to the plume's broader spatial structure. Additionally, anchor nodes are introduced to help UAVs maintain alignment with the plume path as they trace it upwind, enhancing group coordination and reducing the likelihood of losing track of the plume.

\subsubsection{Localize Phase} \textit{Declare} the emitter source. 

In the final phase, UAVs work to localize and declare the the emitter’s estimated position. Operating on a fixed-altitude two-dimensional plane, UAVs continue scanning until either the time limit is reached or a centroid-based termination condition is met. In this condition, UAVs are considered to be ``near'' the source if their collective detection region converges to a consistent centroid estimate. The final estimated emitter location is then defined based on this centroid. Simulation results demonstrate the robustness of this method, particularly under wind turbulence, where traditional strategies may struggle to maintain plume contact or declare accurate source locations.

\section{System Models and Problem Formulation}
\label{sec:ch5-System Model}
\begin{figure}[h!]
   \centering
    \includegraphics[width=0.48\textwidth]{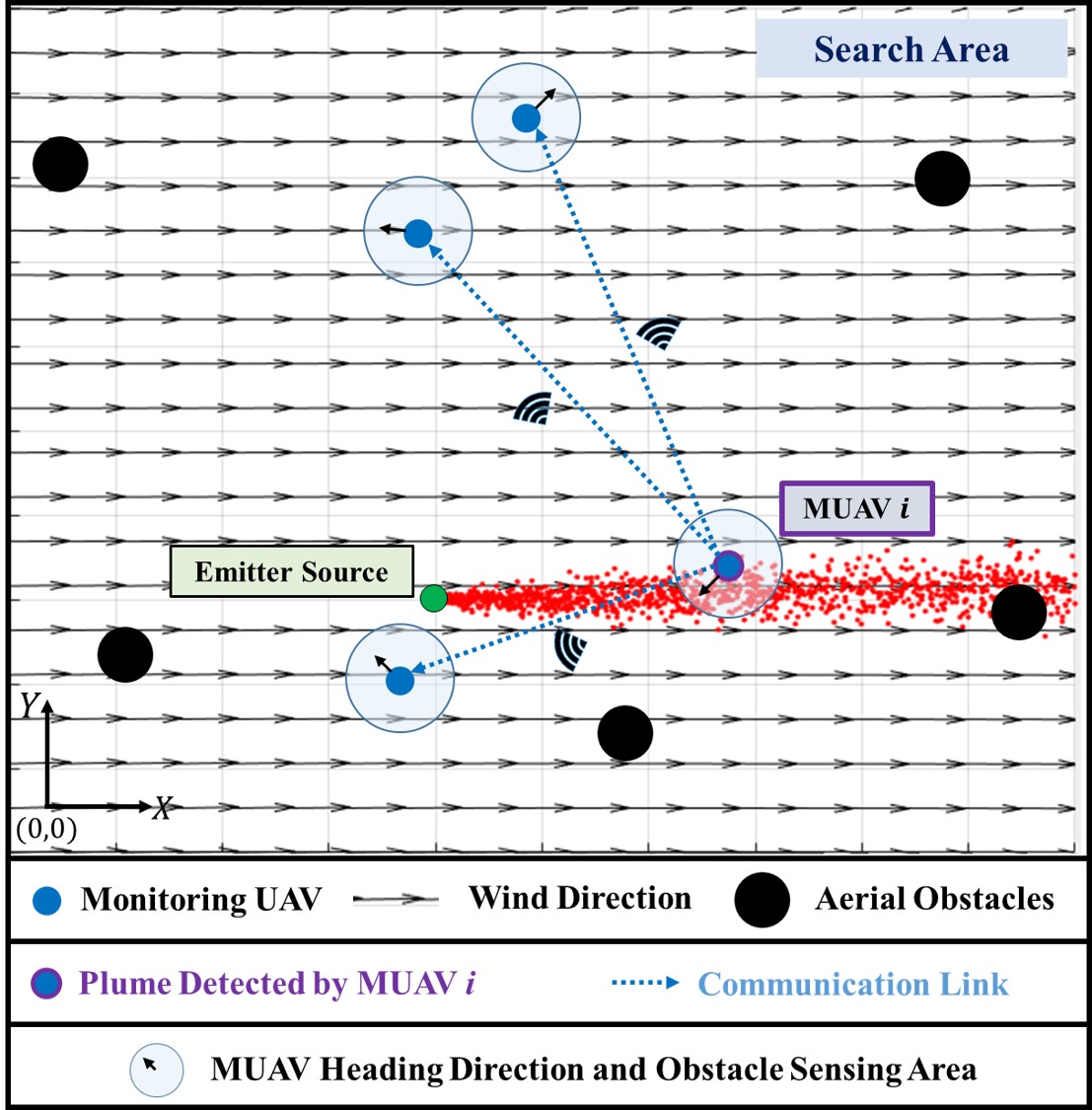}
    \caption{Topology of the considered CPSL Problem.}
    \label{fig:Ch5-TopologyFig}
\end{figure}

We consider a two-dimensional (2D) geographical area of size $X \times Y$ m$^2$ as the search range for a set of $N$ UAVs. The goal of the UAVs is to detect and localize an emitter source. In addition, the UAVs avoid collisions with each other and with a number of aerial obstacles denoted by ($M$), such as birds and other airborne objects. All UAVs operate at the same altitude during the three phases in Sec.~\ref{sec:Ch5-Tasks}. The layout of the 2D topology is illustrated in Fig.~\ref{fig:Ch5-TopologyFig}.

\subsection{On-board Sensors}
Each UAV $i$, for $i \in [1,...,N]$, is equipped with a sensor suite to enable autonomy and successfully accomplishing the CPSL task. The sensor suite is composed of: 
\begin{itemize}
\item A global positioning system (GPS)/inertial navigation system (INS) to monitor the UAV's current position ${\bf p}_{u}^{i}(t) = \left[x_{u}^{i}(t),y_{u}^{i}(t)\right] \in \mathbb{R}^2$ (m) and linear velocity $v_{u}^{i}(t) \in \mathbb{R}$ $\left(\text{m}/\text{s}\right)$ in the inertial frame; 

\item Attitude and gyroscope sensors to monitor orientation information, such as the UAV's heading direction $\theta_{u}^{i}(t) \in \mathbb{R}$ (radians) and angular velocity $\omega_{u}^{i}(t) \in \mathbb{R}$ $\left(\text{rad}/\text{s}\right)$ in the inertial frame; 

\item A chemical sensor to measure the methane concentration ${\tilde{\rho}}_i(t) \in \mathbb{R}$ (ppm); 

\item An anemometer sensor ${\tilde{\bf V}}^w_i (t) = \left[{\tilde{V}}^w_{x,i}(t),{\tilde{V}}^w_{y,i}(t)\right] \in \mathbb{R}^2$ $\left(\text{m}/\text{s}\right)$ to monitor the wind speed and direction in the inertial frame;  

\item Perception/obstacle detection sensors such as radar, to maintain collision avoidance; 

\item A communication and data logger module to enable UAV-to-UAV information exchange.  

\end{itemize}
Here, $t$ represents the time-step index, and the notation $({\tilde{\cdot}})$ indicates measured sensor readings. In this study, we aim to simulate a near-realistic scenario of plume dynamics and associated chemical and wind sensor readings, which are critical for guiding the UAVs toward the emitter location. To reflect field conditions, we assume that only the chemical and wind sensor measurements are affected by noise, accounting for natural fluctuations. The position and attitude information provided by the GPS/INS system and gyroscopes are considered true values without noise. Our focus is on designing a framework to handle transient and sparse measurements from chemical concentration readings and wind dynamics, which are essential for UAV navigation. In future work, we plan to incorporate errors in position, velocity, and acceleration to analyze their impact on localization accuracy.

\begin{figure}[h]
   \centering
    \includegraphics[width=0.35\textwidth]{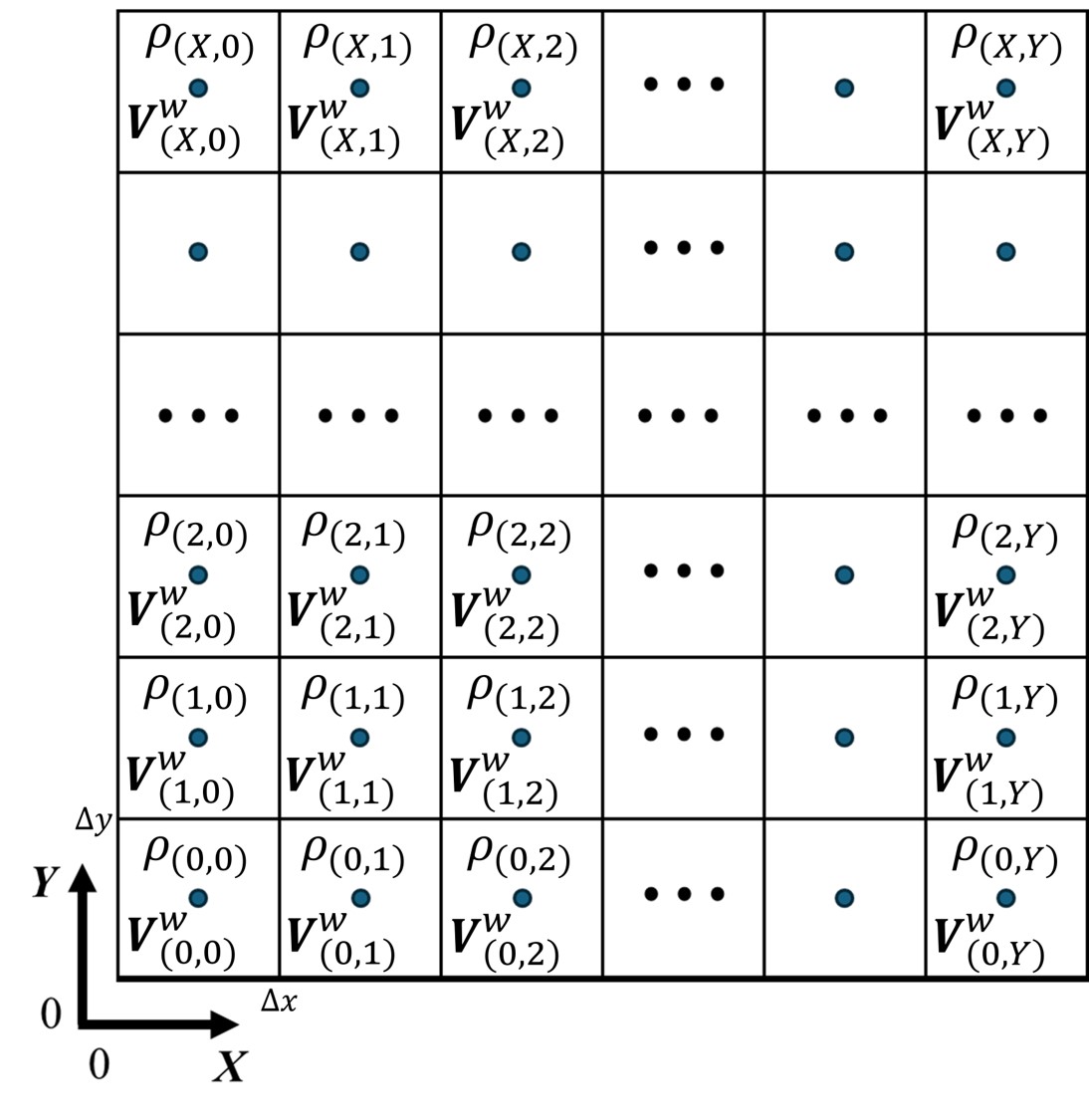}
    \caption{Search area of the defined CPSL problem.}
    \label{fig:Ch5-GridFig}
\end{figure}

During each time-step, the UAVs interact with the environment to gather new information, forming the basis for learning an optimal decision-making strategy. To reduce simulation complexity, the plume concentration and flow field (wind) data of the search area are discretized into cell blocks of size $\Delta x \times \Delta y$ m$^2$, as illustrated in Fig.~\ref{fig:Ch5-GridFig}. This data is organized into a lookup table used during the learning process. For example, if a UAV $i$ is located in cell block $(1,2)$ at time $t$, the corresponding concentration and flow field values it logs are those computed at the center of that block at time $t$: ${\tilde{\rho}}_i(t) = {\tilde{\rho}}_{(1,2)}(t)$ and ${\tilde{\bf V}}^w_i(t) = {\tilde{\bf V}}^w_{(1,2)}(t)$. 

\subsection{Chemical Plume Dynamics}
To simulate the dynamics of a realistic, time-varying plume within a flow field, the UAV's plume measurements should closely mimic those of an experimental sensor. The plume dynamics are based on the filament-based dispersion model presented in \cite{farrell2002filament}, where the plume is generated through the sequential release of ``puffs'' from the emitter source location $({\bf p}_{\mathrm{es}})$. Each puff consists of a number of chemical filaments, which can be further decomposed into ``packets'' of molecules. A dynamic wind model is integrated to simulate the transport of these filaments as the wind direction and speed vary. Specifically, the shape and density of the plume are influenced by wind advection, as well as by turbulent and molecular diffusion. Advection describes the movement of a substance due to the flow of the surrounding velocity field. Turbulent diffusion refers to the mixing and stirring caused by turbulent motions, which can cause some filaments to deviate from the primary flow direction. Molecular diffusion accounts for the movement of particles due to their thermal energy. 

Let us denote the 2D position of the $f$-th filament at time $t$ as ${\bf p}_{p}^{f}(t) = [x_p^{f}(t),y_p^{f}(t)]$, where $f \in [1,F(t)]$ and $F(t)$ is the total number of filaments present at time $t$. The model decomposes the transport of the filaments into two processes: the change in shape of the filament due to molecular and turbulent diffusion ${\hat{\bf V}}^{md}(t)$, and the effect of  wind advection, represented by $\hat{{\bf V}}^w(t)$: 
\begin{align*}
\frac{d{\bf p}_p^{f}(t)}{\,dt} &= {\bf V}_p^{f}(t) = {\hat{\bf V}}^w(t) + {\hat{\bf V}}^{md}(t) \\
\implies {\bf p}_p^{f} (t+1) &= {\bf p}_p^{f} (t) + \Delta t \left({\hat{\bf V}}^w (t) + {\hat{\bf V}}^{md}(t) \right)
\end{align*}
where ${\bf p}_p^{f}(0) = {\bf p}_{\mathrm{es}}$, $\Delta t$ is the time-step length, ${\bf V}_p^{f}(t)$ represents the $(x,y)$ velocity components of the $f$-th filament, and the notation $(\hat{\cdot})$ denotes the numerically solved or true component values. 

\subsubsection{Advection Model}

Let the 2D $(x,y)$ components of the advection term be denoted as ${\hat{\bf V}}^w = [v_1, v_2]$. The advection model is based on the Reynolds-Averaged Navier-Stokes (RANS) equations, which are the time-averaged equations used to describe fluid flow motion. These equations utilize Reynolds decomposition, a mathematical technique that separates a flow variable into its mean (time-averaged) component $(\overline{\cdot})$ and its fluctuating or turbulent component $(')$. Specifically, we have $v_1(x,y,t) = \overline{v_1(x,y)} + {{v'}_1}(x,y,t)$ and $v_2(x,y,t) = \overline{v_2(x,y)} + {{v'}_2}(x,y,t)$. Since the UAVs are confined to a small search area and operate at low altitudes, the local pressure and wind conditions remain relatively constant. Additionally, because we analyze the plume behavior at a filament-based level, we assume that forces due to molecular viscosity, Coriolis effects, and geostrophic winds are negligible compared to turbulent effects.  The time-averaged RANS equations are expressed as: 

\begin{align*}
\frac{\partial {\overline{v_1}}}{\partial t} &= -\overline{v_1}\frac{\partial  {\overline{v_1}}}{\partial x} -\overline{v_2}\frac{\partial  {\overline{v_1}}}{\partial y} -\frac{\partial {\overline{{v'}_1{v'}_1}}}{\partial x} - \frac{\partial {\overline{{v'}_1{v'}_2}}}{\partial y}\\
\frac{\partial {\overline{v_2}}}{\partial t} &= -\overline{v_1}\frac{\partial  {\overline{v_2}}}{\partial x} -\overline{v_2}\frac{\partial  {\overline{v_2}}}{\partial y} -\frac{\partial {\overline{{v'}_2{v'}_1}}}{\partial x} - \frac{\partial {\overline{{v'}_2{v'}_2}}}{\partial y}
\end{align*}
which satisfy the continuity equation: 
\begin{align*}
0 = \frac{\partial  {\overline{v_1}}}{\partial x} + \frac{\partial  {\overline{v_2}}}{\partial y}. 
\end{align*}
The incompressible Navier-Stokes equations can be further simplified by expressing the fluctuating terms in terms of the time-averaged components  \cite{farrell2002filament}: 
\begin{align*}
\frac{\partial {\overline{v_1}}}{\partial t} &= -\overline{v_1}\frac{\partial  {\overline{v_1}}}{\partial x} -\overline{v_2}\frac{\partial  {\overline{v_1}}}{\partial y} + \frac{1}{2}K_x\frac{\partial ^ 2 {\overline{v_1}}}{\partial x^2} + \frac{1}{2}K_y \frac{\partial ^ 2 {\overline{{v_1}}}}{\partial y^2}, \\ 
\frac{\partial {\overline{v_2}}}{\partial t} &= -\overline{v_1}\frac{\partial  {\overline{v_2}}}{\partial x} -\overline{v_2}\frac{\partial  {\overline{v_2}}}{\partial y} + \frac{1}{2}K_x\frac{\partial ^ 2 {\overline{v_2}}}{\partial x^2} + \frac{1}{2}K_y \frac{\partial ^ 2 {\overline{{v_2}}}}{\partial y^2},
\end{align*}
where $K_x$ and $K_y$ are diffusivity coefficients. These equations can be solved numerically using finite difference methods at the grid points of the search area, with interpolation between grid points, subject to specific boundary conditions  (refer to Sec.~2 in \cite{hollenbeck2022digital}). 

The meandering of the flow field from the mean flow is modeled using a colored noise process  \cite{farrell2002filament}. A white Gaussian noise process with zero mean and unit variance is passed through a filter with the following transfer function:  
\begin{align*}
H(s) = G\frac{a}{s^2 + bs + a}
\end{align*}
where $G$ is the gain of the filter, and $a$ and $b$ are the damping ratio and natural frequency of the filter, respectively. We vary the value of $G$ in Sec.~\ref{sec:Ch5-NumericalResults} to evaluate system performance under different levels of plume meandering.

\subsubsection{Diffusion Model} 
Let the 2D $(x,y)$ components of the relative diffusion term be denoted as ${\bf V}^{md} = [v_{m_1},v_{m_2}]$. The model is characterized by a zero-mean white noise process, which reflects the stochastic turbulence of the molecules.  The spectral density of this noise is  $\sigma_{v_{m,1}}$ and $\sigma_{v_{m,2}}$ in units of m/s/$\sqrt{\text{Hz}}$ \cite{farrell2002filament}.\\

\noindent {\textbf{SUMMARY:}} By adjusting parameters such as the mean flow, colored noise characteristics, and variance levels, different flow fields and corresponding plume trajectories can be simulated.  The instantaneous wind measurement recorded by UAV $i$ at the point ${\bf p}_{u}^{i}(t) = [x_{u}^{i}(t),y_{u}^{i}(t)]$ (which is within some cell block $(a,b)$) is decomposed into the contributions from advection, diffusion and sensor noise $({\bf V}^{w}_{n_i}(t))$: 
\begin{align}
{\tilde {\bf V}}^{w}_i(t) = {\hat {\bf V}}^{w}_{i}(t) + {\bf V}^{w}_{n_i}(t).
\end{align}
Here, ${\hat {\bf V}}^{w}_{i}(t)$ represents the true or  numerically solved velocity component values, while ${\bf V}^{w}_{n_i}(t)$ is characterized by white Gaussian noise with a variance of $\sigma_{V_n}^2$. The wind speed components in ${\tilde{\bf V}}^w_i(t)$ are the values stored in UAVs' data logger. 

\subsection{Concentration Fluctuation Model}

The instantaneous point concentration measurement recorded by UAV $i$ at position ${\bf p}_{u}^{i}(t) = [x_{u}^{i}(t),y_{u}^{i}(t)]$ (which lies within cell block $(a,b)$) is decomposed into contributions from the filaments and the noise present in the sensor $(\rho_{n_i}(t))$: 
\begin{align}
{\tilde \rho}_i(t) = {\hat \rho_i(t)} + \rho_{n_i}(t) 
\end{align}
where $\hat{\rho}_i(t)$ is the filtered total concentration from the filaments present at time $t$, and $\rho_{n_i}(t)$ is characterized by white Gaussian noise with variance $\sigma_{\rho_n}^2$ and a background CH$_4$ noise bias $b_{\rho_n}$. ${\tilde {\rho}}_i(t)$ is the concentration stored in UAVs' data logger. 

The total concentration contribution from the filaments within cell block $(a,b)$ at time $t$ is given by 
\begin{align}
\rho_i(t) = \rho_{(a,b)}(t) = \sum\limits_{f = 1}^{F(t)}C_{f}(t)\;\; [\text{ppm}]
\end{align}

\begin{align}
&C_f(t) = \frac{Q}{\sqrt{8 \pi ^3}R_{f}^3(t)} \exp\left( \frac{-||{\bf p}_{(a,b)} - {\bf p}_{u}^{f}(t)||^2}{R_f^2(t)}\right)\;\;\; \frac{\text{molecules}}{\text{cm}^3}\\
&C_f(t)\;\;\; [\text{ppm}]= \frac{C_f(t) \times 10^6}{N_0} \nonumber 
\end{align}
where 
\begin{itemize}
\item $F(t)$ is the total number of filaments present in the search area at time $t$.

\item ${\bf p}_{(a,b)} = [x_{a},y_{b}]$ is the center point of cell block $(a,b)$.

\item ${\bf p}_{u}^{f}(t) = [x_{u}^{f}(t),y_{u}^{f}(t)]$ is the 2D position of the $f$-th filament at time $t$. 

\item $N_0 = \frac{Pk}{TR}$ is the density of methane gas where $k$ is Avogadro's number, $R$ is the molar gas constant, $P$ is the ambient pressure, and $T$ is the ambient temperature.

\item The growth of the $f$-th filament is modeled in terms of its radius $R_f$. The change in radius as a function of time is given by \cite{farrell2002filament}
\begin{align*}
R_f(t) = R_f(t-1) + \gamma \Delta t,\; t > 0 
\end{align*}
where $\gamma > 0$ is a constant, and $R_f(0)$ is the initial size of the filament when it emerges from the emitter location. 

\item $Q = \frac{\overline{Q}}{N_f}$ where $\overline{Q}$ is the molecule release rate in $\frac{molecules}{sec.}$ and $N_f$ is the filament release rate in $\frac{filaments}{sec.}$.
\end{itemize}

\begin{figure}[h!]
   \centering
    \includegraphics[width=0.49\textwidth]{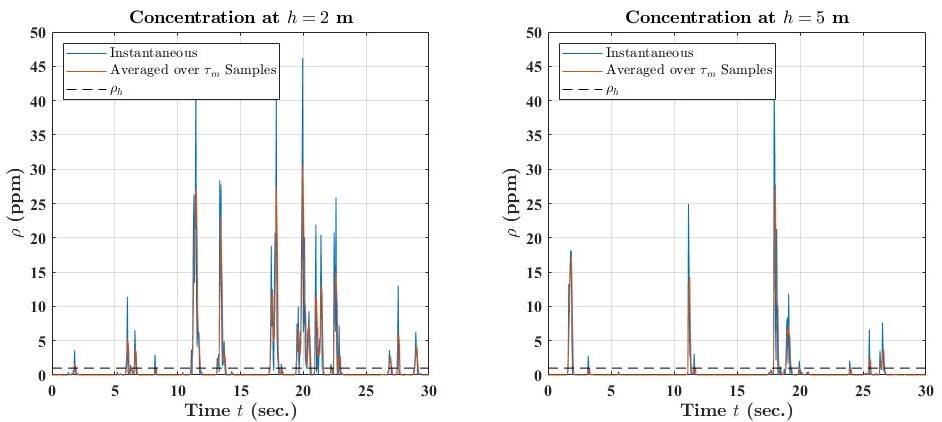}
    \caption{Concentration plot when a UAV is at $h = 2$ m (left) and $h = 5$ m (right) at a point 6 m downwind from the emitter location along the plume's center-line. The time resolution is 0.05 sec. and the filament release rate is $N_f = 50$ filaments/sec.}
    \label{fig:Ch5-PlumeTimePlot}
\end{figure}

The concentration sensed by the CH$_4$ sensor is modeled as the response of a simple low-pass filter with bandwidth $B$, followed by a threshold detector with threshold $c_h$: 
\begin{align}
\rho_i(t) &= \rho_i(t) + B\Delta t \left(\rho_i(t) - {\hat{\rho}}_i(t-1) \right)\\
\hat{\rho}_i(t) &= \begin{cases}
\rho_i(t) & \text{if } \rho_i(t) > c_h\\
0 & o.w.
\end{cases}
\end{align}
where $\hat{\rho}_i(t)$ is the filtered output concentration of the sensor (excluding noise), and $\rho_i(t)$ is the instantaneous concentration input to the filter. Fig. \ref{fig:Ch5-PlumeTimePlot} illustrates the instantaneous and averaged sensed concentration values at two different altitudes of the simulated plume in Example 1 at a fixed point in the environment grid. The dashed line indicates a cutoff threshold ($\rho_{\mathrm{th}} = 1$ ppm), suggesting that the UAV has detected a significant chemical signature and may be within the plume. From the plots, we can observe the sparsity and short duration of these pulses, even when relatively close to the emitter source.  At an altitude of $h = 5$ m, significant gaps between pulses suggest that there may not be enough chemical content present at this location due to the wind profile.  Notice the large peak at $t \approx 17$ seconds for $h = 2$, which may indicate a recently released puff of filaments. Similarly, the large peak at $t \approx 18 $ seconds for $h = 5$ m shows that filaments transported upward by the wind profile produced a similar chemical reading at $5$ m. This variability in chemical concentrations at different altitudes presents a key challenge in designing an algorithm capable of efficiently tracing a plume based on inherently sparse chemical signatures. This concentration fluctuation model closely aligns with the experimental results shown in Fig.~4 of \cite{morales2022controlled}. A significant research question addressed in this work is : \textit{How effectively can we identify the emitter source based on the limited sensor information available? } 

\subsubsection{\textbf{Example 1: Filament Dispersion From Emitter}}

\begin{figure}[h!]
   \centering
    \includegraphics[width=0.4\textwidth]{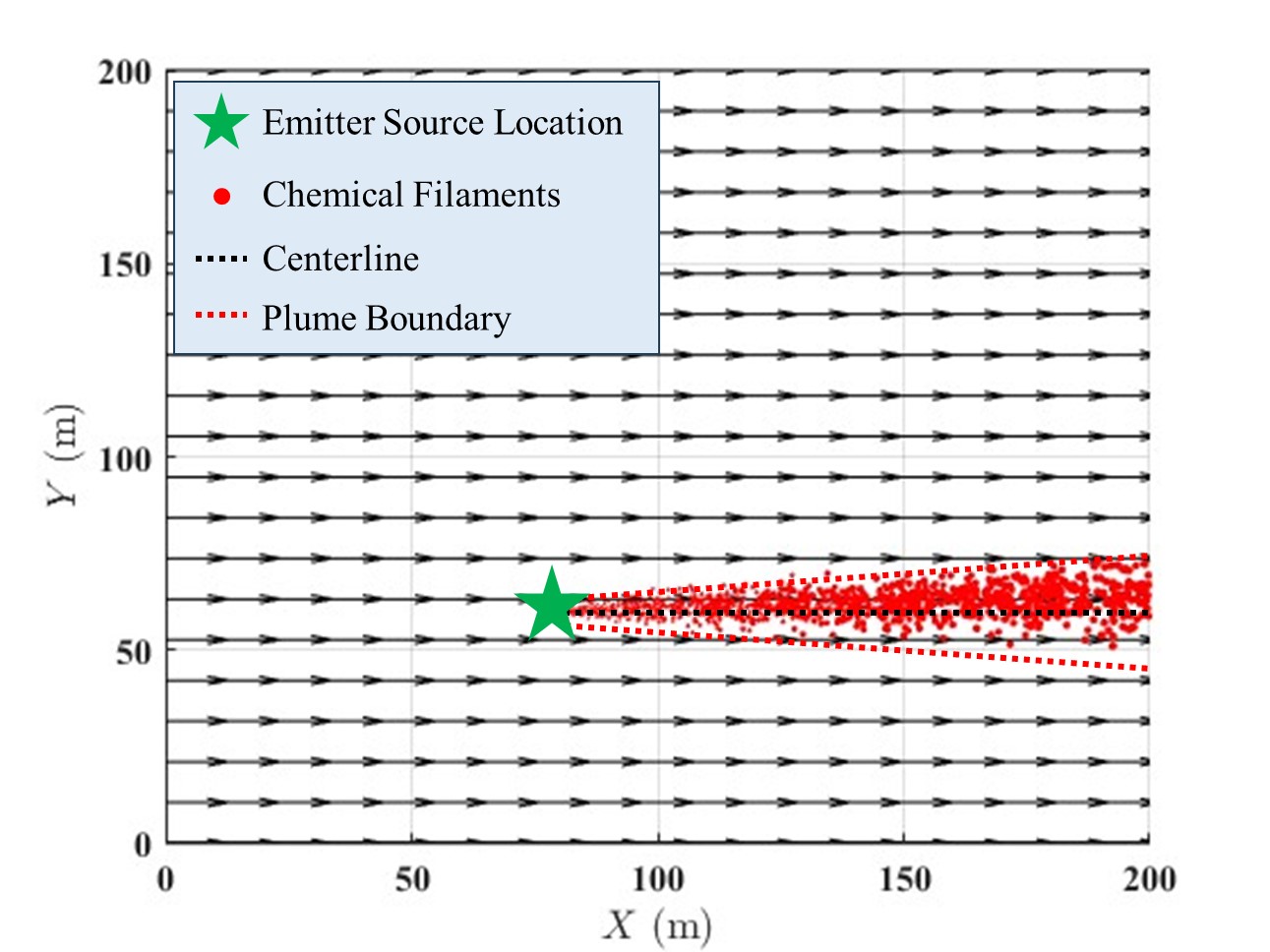}
    \caption{Plume dynamics due to relative diffusion and constant advection, i.e. the mean flow field has velocity ${\overline{\bf V}}^w = [1,0]$ m/s, and the emitter source location is at $(x,y) = (80, 60)$ m.}
    \label{fig:Ch5-ExPlume}
\end{figure}

Fig. \ref{fig:Ch5-ExPlume} demonstrates a top-down view of a simulated plume\footnote{Animation of this plume example is posted on YouTube: \url{https://www.youtube.com/watch?v=_qtImvQBi4w}} showing the advective and diffusive behavior of the filaments within a 200 x 200 m grid. The emitter location, indicated by a green star, serves as the source of the plume. The simulation parameters are listed in Table \ref{table:Ch5-PlumeParameterTable}. The equations described earlier are numerically solved with a mean advection flow (${\overline{\bf V}^w} = [1,0]$ m/s) in the $+X$ direction, indicating no meandering of the centerline due to the constant wind velocity. The simulation incorporates the effects of colored noise and relative diffusion. In the figure, arrows represent the instantaneous direction of the flow field.  The filament density is higher near the emitter location, showing the initial concentration of the release. As distance from the emitter increases, the radius of the filaments grows, causing the density to become more sparse as the filaments are carried away by the wind. The plume's centerline is shaped by the wind experienced by each filament, forming a coherent plume structure. 

\begin{table}[h!]
\centering
\caption{Plume Parameter Values}
\label{table:Ch5-PlumeParameterTable}
\begin{tabular}{|p{4.2cm}|c|}
\hline
\textbf{Notation} & \textbf{Value}\\
\hline 
Emitter Source Location $\left({\bf p}_{es}\right)$ & (80, 60) m\\
\hline 
Mean Direction of Wind $\left({\overline{\bf V}^w}\right)$ & [1,0] m/s\\
\hline 
Colored Noise Constants $\left([a,b,G]\right)$& $[0.005,\; 0.02,\; 1]$\\ 
\hline 
Concentration Sensor Filter Coeffs.\newline $\left(B,c_h\right)$ & $(0.1,10^{-4})$\\ 
\hline 
Diffusivity Constants $\left( K_x,K_y \right)$ & (1000,1000) $\text{m}^2$/s \\ 
\hline 
Relative Diffusion Spectral Density\newline $\left( \sigma_{v_{m,1}}, \sigma_{v_{m,2}}\right)$& $(2,2)$ m/s/$\sqrt{\text{Hz}}$ \\ 
\hline 
Initial Radius of Filaments $\left(R_f(0)\right)$ & 0.01 cm\\
\hline 
CH$_4$ Sensor Noise Bias $\left(b_{\rho_n}\right)$ & 1.98 ppm\\
\hline 
CH$_4$ Sensor Noise Variance $\left(\sigma_{\rho_n}^2\right)$ & 0.005\\
\hline 
Anemometer Sensor Noise Variance\newline $\left(\sigma_{V_n}^2\right)$ & 0.01\\
\hline 
Ambient Pressure $\left(P\right)$ & 101,325 Pa\\
\hline 
Avogadro's Number $\left(k\right)$ & $6.02214076 \times 10^{23}$ 1/mol.\\
\hline 
Ambient Temperature $\left(T\right)$ & 288 K\\
\hline 
CH$_4$ Molar Gas Constant $\left(R\right)$ & 8.31446 J/(mol.$\times$K)\\
\hline 
Filament Growth Rate $\left(\gamma\right)$& 0.001 cm/s\\
\hline 
Filament Release Rate $\left(N_f\right)$ & 50 filaments/s\\
\hline 
Molecule Release Rate $\left(\overline{Q}\right)$ & $1.967243976 \times 10^{21}$ $\frac{\text{molecules}}{\text{filament}}$\\
\hline 
\end{tabular}
\end{table}

\subsection{UAV Dynamics}
\label{Ch5-UAV-Dynamics}
\subsubsection{UAV Motion Model}
In the UAV dynamics model, the altitude of the UAVs is restricted between two values, $h_{min}$ and $h_{max}$. Here, we consider a simple planar model to describe the change in the position and heading direction of a UAV $i$: 

\begin{alignat}{2}
\label{eqn:Ch5-system-eqn}
\begin{bmatrix}
{\dot{\bf p}_{u}^{i}(t)} \\
{\dot{\theta}_{u}^{i}(t)}
\end{bmatrix} = \begin{bmatrix}
{\dot{x}_{u}^{i}(t)} \\
{\dot{y}_{u}^{i}(t)} \\ 
{\dot{\theta}_{u}^{i}(t)}
\end{bmatrix} = &\begin{bmatrix}
\cos(\theta_{u}^{i}(t))    & 0\\
\sin(\theta_{u}^{i}(t))   & 0\\
0 & 1\\ 
\end{bmatrix}&&\begin{bmatrix}
v_{u}^{i}(t)\\
\omega_{u}^{i}(t)
\end{bmatrix},\;\;\; i \in [1,...,N].\\
&{\;\;\;\;\;\;\;\;\text{System}} && {\;\;\text{Control}} \nonumber \\ 
&{\;\;\;\;\;\;\;\;\text{Matrix}} && {\;\;\text{Vector}} \nonumber 
\end{alignat}
In this model:
\begin{itemize}
    \item The UAV's linear speed is bounded by $ v_{u}^{i}(t) \in \left[v_{min},v_{max}\right]$ m/s.
    \item  The UAV's angular speed is bounded by $\omega_{u}^{i}(t) \in \left[\omega_{min},\omega_{max}\right]$ rad/s. 
\end{itemize}

Additionally, there are $M$ aerial obstacles moving within the search area. The position of obstacle $m$ at time $t$ is denoted as ${\bf p}_o^m(t) = \left[x_o^m(t),y_o^m(t) \right] \in \mathbb{R}^2$ (meters), for $m \in [1,...,M]$. We assume that all obstacles move throughout the search area with a linear velocity $v_o^m(t) \in \mathbb{R}$ m/s and heading angle $\theta_o^m(t) \in \mathbb{R}$ (radians). 

\subsubsection{UAV Collision Model}
All UAVs and obstacles are modeled as circular shapes with radii $r_{u}$ and $r_{o}$ (in meters), respectively. A collision between two UAVs occurs if the following condition is met: 
\begin{align}
d_{u}^{i,i'}(t) = ||{\bf p}_u^{i}(t) - {\bf p}_{u}^{i'}(t)|| < 2r_{u} + \varepsilon 
\end{align}
where $d_{u}^{i,i'}(t)$ is the relative distance between UAVs $i$ and $i'$, $\varepsilon$ is a safety distance coefficient, and $||\cdot ||$ denotes the Euclidean norm. A collision occurs between UAV $i$ and obstacle $m$ if the following condition is met: 
\begin{align}
d_{u,o}^{i,m}(t) = ||{\bf p}_u^{i}(t) - {\bf p}_{o}^{m}(t)|| < r_{u} + r_{o} + \varepsilon,  
\end{align}
where $d_{u,o}^{i,m}(t)$ is the relative distance between UAV $i$ and obstacle $m$. 
To ensure safety, the UAVs follow a collision safety protocol \cite{pierre2023multi} that overrides any actions leading to potential collisions or that cause the UAV to go beyond the boundaries of the simulated environment. 

\subsection{Objective of the UAV Agents}
\label{sec:Ch5-Objective}
Upon detecting the chemical plume, the UAVs establish a virtual anchor point (more details in Sec.~\ref{sec:Ch5-DRL-Design}) to guide their collective movement. Instead of each UAV independently tracing the plume, the anchor point serves as the shared target for coordination. UAVs measure chemical concentrations in the vicinity of the anchor and contribute to updating its position in the upwind direction---when a predefined threshold condition ($\overline{\tilde{\rho}}_i(t) > \rho_\mathrm{th}$) is met. In this design, the anchor point, rather than the UAVs themselves, acts as the primary tracer. This approach enhances robustness, especially when UAVs are near the emitter, as it reduces the risk of overshooting the source by maintaining stability of the anchor.

The UAV agents are assumed to have no prior knowledge of the emitter’s location. Their understanding of the plume’s behavior is acquired solely through local sensor measurements and shared information from neighboring agents. Each UAV’s objective is to control its linear and angular velocities, as defined in the system dynamics in Eqn.~(\ref{eqn:Ch5-system-eqn}), to achieve three main goals: (1) explore the chemical gradient or signature around the anchor point, (2) contribute to the consistent upwind advancement of the anchor, and (3) avoid collisions with other UAVs and obstacles.

\section{Multi-Agent Deep Reinforcement Learning for UAV Formation Control Optimization in CPSL}
\label{sec:Ch5-MADRL}

The multi-UAV control for CPSL is modeled as a multi-agent DRL problem, utilizing the CTDE paradigm, and framed within the context of a partially observable Markov decision process (POMDP), as outlined in our previous work \cite{sherman2024counter}. By definition, a MDP represents a sequential decision-making framework where agents learn by interacting with the environment. In line with our earlier research \cite{sherman2024counter}, each agent operates in a state $s$, selects an action $a$ based on a policy $\pi(a|s)$, transitions to a new state $s'$, and receives a reward $r$ at the subsequent time step $t+1$. 
This process continues until one of the following termination conditions is met: (1) the UAVs declare the emitter has been located, or (2) the episode duration reaches its predefined time limit. In the following sections, we will outline the specific definitions of the reinforcement learning elements, the anchor update rule, and the episode termination condition.


\subsection{Reinforcement Learning (RL) Framework}
\label{sec:Ch5-DRL-Design}


\subsubsection{State Space}
\begin{figure}[h]
   \centering
    \includegraphics[width=0.49\textwidth]{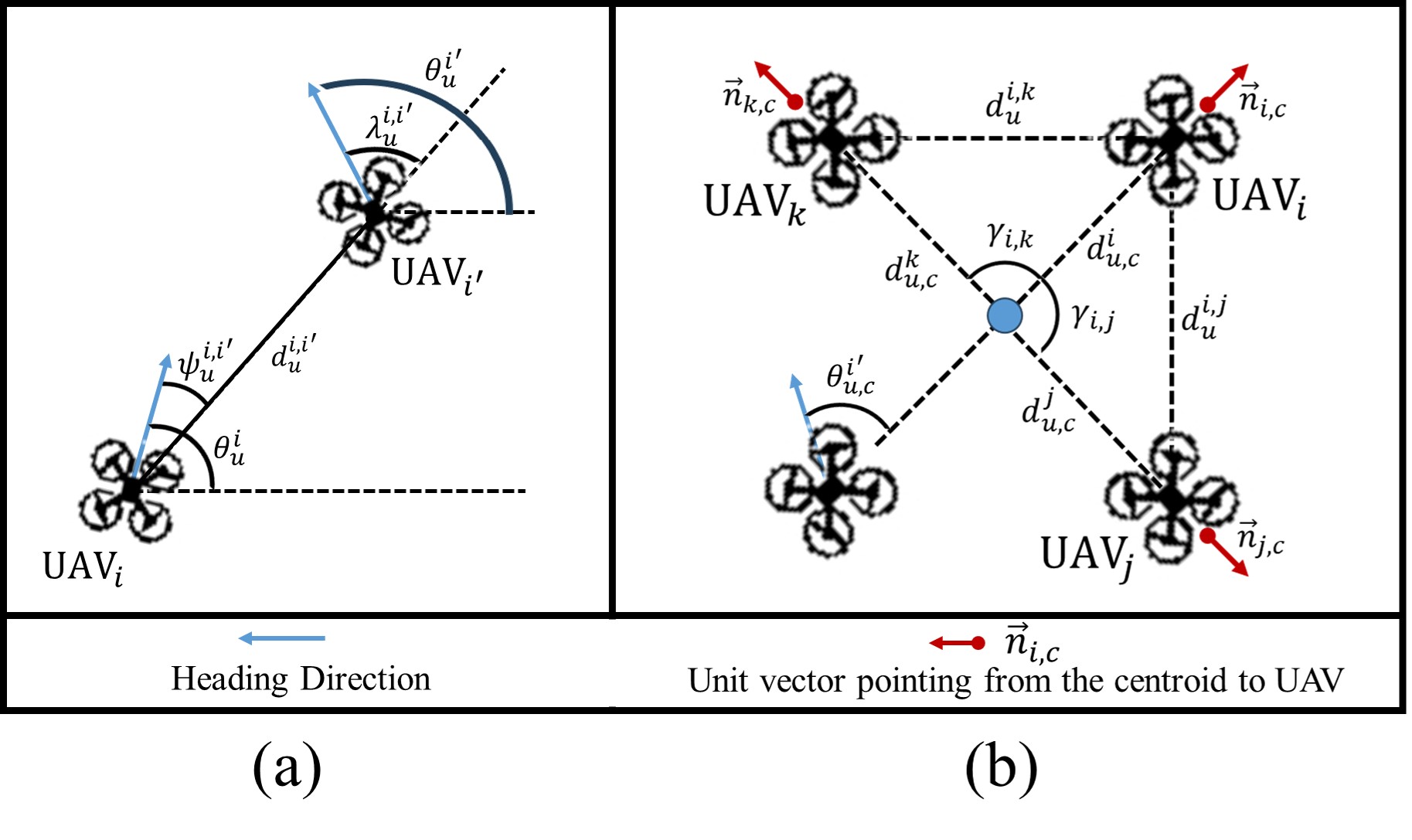}
    \caption{(a) Relative angle information of UAV $i'$ observed by UAV $i$ (b) relative formation information observed by UAV $i$.}
    \label{fig:Ch5-RelativeDefinitions}
\end{figure}

Let $\mathcal{S}$ denote the state space of the UAVs. The state of UAV $i$ at time-step $t$, denoted as ${\bf s}_u^{i}(t) \in \mathcal{S}$, is composed of the following elements: 
\begin{align}
{\bf s}_u^{i}(t) = \left[{\bf p}_u^i(t),\theta_u^i(t),v_u^i(t),\omega_u^i(t) \right], 
\end{align}
where 
\begin{itemize}
\item ${\bf p}_u^i(t) = \left[x_u^i(t),y_u^i(t)\right]$ represents the 2D position coordinates of UAV $i$ at time $t$, 
\item $\theta_u^i(t)$ is the heading angle of UAV $i$ at time $t$ (depicted in Fig. \ref{fig:Ch5-RelativeDefinitions}(a)),
\item $v_u^i(t)$ is the linear velocity of UAV $i$ at time $t$, and 
\item $\omega_u^i(t)$ is the angular velocity of UAV $i$ at time $t$. 
\end{itemize}

\subsubsection{Observation Space}
Let $\mathcal{O}$ denote the observation space of the UAVs. The total observation vector of UAV $i$ at time-step $t$, denoted as ${\bf o}_u^i(t) \in \mathcal{O}$, is composed of the following elements: 

\begin{align}
{\bf o}_{u}^{i}(t) = \left[{\bf o}_{u,s}^i(t),{\bf o}_{u,c}^i(t),\{{\bf o}_{u}^{i,i'} (t)\}, \{{\bf o}_{u,s}^{i,i'} (t)\}, {\bf o}_{a}^i(t), \{{\bf o}_{u,o}^{i,m} (t) \}\right]
\end{align}
for $i' \in [1,N]{\backslash}i$ and $m \in [1,M]$. 

\begin{enumerate}[label=\arabic*.,leftmargin=*,listparindent=1.5em] 
    \item UAV Sensor Measurements $\left( {\bf o}_{u,s}^i(t) \right)$
    
    Due to the noise and short time-scale in concentration and flow field readings, a moving average filter of length $\tau_m$ is applied. This filter averages $\tau_m$ consecutive samples of the concentration and flow field measurements, storing the result in the sensor observation vector: 

    \begin{align}
    {\bf o}_{u,s}^i(t)=\left[\overline{\tilde{\rho}}_i (t),\overline{\tilde{\bf V}}^w_i(t), q_i(t),h_i(t)\right], 
    \end{align}
    where 
    \begin{itemize}
    \item $\overline{\tilde{\rho}}_i (t) = \frac{1}{\tau_m}\sum\limits_{t' = t - \tau_m}^{t} \tilde{\rho}_i(t')$, for $t' \in [t - \tau_m,t]$, is the averaged point concentration measured by UAV $i$ over the past $\tau_m$ samples,   
    \item $\overline{\tilde{\bf V}}^w_i (t) = \left[\overline{{\tilde{V}}}^w_{x,i}(t), \overline{\tilde{V}}^w_{y,i}(t)\right]$ holds the averaged 2D flow field coordinates measured by UAV $i$ where 
    \begin{align*}
    \overline{\tilde{V}}^w_{x,i}(t) = \frac{1}{\tau_m}\sum\limits_{t' = t - \tau_m}^{t} \tilde{V}^w_{x,i}(t') \;\; \text{and}\;\; \overline{\tilde{V}}^w_{y,i}(t) = \frac{1}{\tau_m}\sum\limits_{t' = t - \tau_m}^{t} \tilde{V}^w_{y,i}(t'), 
    \end{align*}
    \item $q_i(t)$ is an indicator variable for plume detection, where 
    \begin{align*}
    q_i(t) = \begin{cases}
    1, & \text{if } \overline{\tilde{\rho}}_i(t) - 1.98 \geq \rho_{\mathrm{th}}\\
    0, & o.w.
    \end{cases}
    \end{align*}
    This threshold accounts for sensor noise and the background bias level. 

    \item $h_i(t)$ represents the current altitude of UAV $i$ at time $t$. 
    
    \end{itemize}
    
    \item Relative Information to the Formation Centroid $\left({\bf o}_{u,c}^i(t) \right)$

    This observation tracks UAV $i$'s relative position to the formation centroid (depicted in Fig. \ref{fig:Ch5-RelativeDefinitions}(b)): 
    
    \begin{align}
    {\bf o}_{u,c}^i (t)= \left[d_{u,c}^i(t), \theta_{u,c}^i(t)\right]
    \end{align}
    where 
    \begin{itemize}
    \item $d_{u,c}^i(t)$ is the distance between UAV $i$ and the centroid of the formation, and 
    \item $\theta_{u,c}^i(t)$ is the angle formed between the heading of UAV $i$ and the position of the centroid.
    \end{itemize}

   The ground control station (GCS) calculates this relative information to maintain formation as UAVs trace the plume. The formation size is determined by the number of UAVs and safety margins for collision avoidance.
    
    \item State of Other Agents $\left(\{{\bf o}_{u}^{i,i'} (t)\}\right)$
    
    This vector monitors the states of UAV $i'$ relative to UAV $i$ (depicted in Fig. \ref{fig:Ch5-RelativeDefinitions}(a)): 
    \begin{align}
    {\bf o}_{u}^{i,i'}(t) = \left[d_{u}^{i,i'}(t),\psi_{u}^{i,i'}(t),\lambda_{u}^{i,i'}(t)\right],\;\; \text{for } i' \in [1,N]{\backslash}i,
    \end{align}
    where 
    \begin{itemize}
    \item $d_{u}^{i,i'}(t)$ is the distance between UAV $i$ and UAV $i'$,
    \item $\psi_{u}^{i,i'}(t)$ is the angle between UAV $i$ and UAV $i'$,
    \item $\lambda_{u}^{i,i'}(t)$ is the relative angle between UAV $i$ and UAV $i'$ from UAV $i'$'s frame of reference.
    \end{itemize}

    This information is used for collision avoidance when another UAV enters the sensing radius. 

    \item Sensor Measurements from Other UAVs $\left(\{{\bf o}_{u,s}^{i,i'} (t)\}\right)$
    
    This vector allows UAVs to share sensor data: 

    \begin{align}
    {\bf o}_{u,s}^{i,i'} (t)=\left[\overline{\tilde{\rho}}_{i'} (t),\overline{\tilde{\bf V}}^w_{i'}(t),q_{i'}(t),h_{i'}(t)\right],\;\; \text{for } i' \in [1,N]{\backslash}i,
    \end{align}
    where there elements represent the same quantities as in UAV $i$'s sensor vector but for UAV $i'$.
    
    In future work, we plan to incorporate a latency-aware architecture \cite{sherman2024counter} to evaluate the impact of communication delays in sensor data exchange between UAVs. 

    \item State of Anchor Node $\left({\bf o}_{a}^i(t)\right)$

   Upon detecting the chemical plume, an anchor node is established at the point of detection. Each UAV then records its current relative information to this anchor node, allowing it to adjust its position to move closer to the plume. At each time step, the averaged concentration measurements from the UAVs are compared to a predefined threshold value ($\rho_{\mathrm{th}}$). This value depends on the sensor noise and background bias level and can be determined through the seek phase. If the concentration for a UAV $i$ is higher than $\rho_{\mathrm{th}}$, a new anchor node is established at time-step $t$. The observation vector for the anchor node is given by: 

    \begin{align}
    {\bf o}_{a}^{i}(t) = \left[d_{a}^{i}(t),\theta_{a}^{i}(t), h_{a}(t)\right],
    \end{align}
    where 
    \begin{itemize}
    \item $d_{a}^{i}(t) = ||{\bf p}_{u}^{i}(t) - {\bf p}_{a}(t)||$ is the relative distance between UAV $i$ and the anchor node at position ${\bf p}_{a}(t) = \left[x_a(t),y_a(t) \right]$; 
    \item $\theta_{a}^{i}(t)$ is the relative bearing angle between UAV $i$'s heading and the anchor node; and 
    \item $h_{a}(t)$
    is altitude at which the anchor node was established.
    \end{itemize}

    This observation vector is populated only when an anchor node is established. The anchor node can only be updated if the following conditions are met: 

    \begin{itemize}[label={}]
    \item {\bf (1)} The concentration at the potential new anchor, $\overline{\tilde{\rho}}_i(t)$, is higher than the threshold $\rho_{\mathrm{th}}$, which indicates that the UAV is located within the plume. The initial anchor node is established as soon as a UAV detects the plume (i.e. when $\overline{\tilde{\rho}}_i(t) > \rho_{\mathrm{th}}$ for some $i$).

    \item {\bf (2)} The new anchor node is located upwind of the previous anchor node. 
    \end{itemize} 
    
    The psuedocode in Algorithm \ref{anchor-update} describes the process for updating the anchor node. If detection occurs, the UAV with the highest concentration reading $(\overline{\tilde{\rho}}_{i^{*}}(t))$ is identified, and this value is compared to a predefined concentration threshold $(\rho_{\mathrm{th}})$. If the concentration is higher than the threshold, the algorithm calculates the vector from the previous anchor to UAV $i^{*}$ ($\Delta {\bf p}_a(t)$, as depicted in Fig.~\ref{fig:Ch5-anchor-angle}) and the mean wind direction (${\bf V}$). The angle between the opposite wind direction $-{\bf V}$ and $\Delta {\bf p}_a(t)$ ($\beta$ in Fig.~\ref{fig:Ch5-anchor-angle}) is computed. If $\beta$ falls within a specified field-of-view boundary ($\beta_{max}$), the new anchor is considered upwind, and the anchor information is updated.

\begin{figure}[h]
   \centering
    \includegraphics[width=0.48\textwidth]{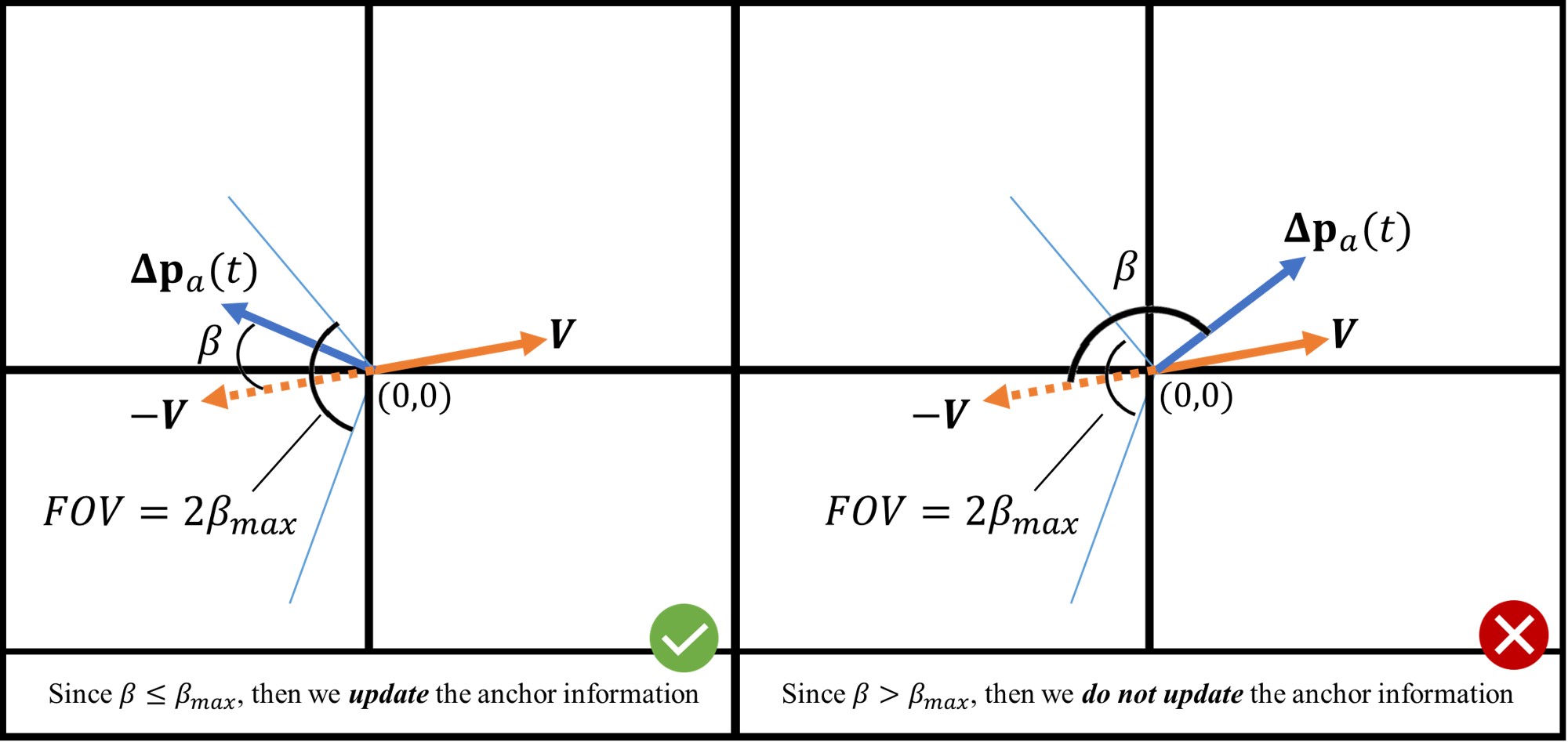}
    \caption{(a) Example of conditions where the anchor information is updated and (b) Example of conditions where the anchor information is NOT updated.}
    \label{fig:Ch5-anchor-angle}
\end{figure}

    Why is this update necessary? Why not simply update the anchor based on concentration alone? Traditional methods often prioritize guiding the formation toward locations with the highest concentration readings. However, this approach can mislead the UAVs, directing them away from the emitter or downwind of the plume. For instance, at a particular moment, the concentration detected downwind may be higher than that detected upwind, leading the formation in the wrong direction, away from the emitter. 
    
    To prevent the UAVs from converging on suboptimal locations or following misleading data, anchor nodes help efficiently trace the plume. The GCS can store and monitor anchor data, offloading processing and storage from the UAVs. UAVs only need to monitor their distance, heading, and altitude relative to the anchor. Anchor updates must take into account both wind and concentration data.
    
    \begin{algorithm}
    \caption{Anchor Node Update Rule}
    \label{anchor-update}
        {\eIf{$\exists \; q_{i}(t) = 1$, \text{ for } $i \in [1,...,N]$}
        {
            Define $\{J\} \gets$ Set of agents where $q_{i}(t) = 1$\;  
            $\{i^{*}\}\gets \argmax\limits_{i, \text{ for }i \in \{J\}} \left(\overline{\tilde{\rho}}_i(t) - 1.98 \right)$ \\ 
            \quad\quad\quad (grab agent(s) with highest detection); \\ 
            {\If {length$\left(\{i^{*}\}\right) > 1$}
            {
            $i^{*}\gets$ Pick the first agent in set $\{i^{*}\}$\; 
            }
            }
            {\eIf {${\overline{\tilde{\rho}}_{i^{*}}(t)} > \rho_{\mathrm{th}}$}
            {{\eIf {$\rho_a(t-1) = 0$}
            {
            $\rho_a(t) = \overline{\tilde{\rho}}_{i^{*}}(t) - 1.98$; 
            ${\bf p}_a(t) = {\bf p}_{i^{*}}(t) $; 
            $h_a(t) = h_{i^{*}}(t)$; 
            }
            {
            \text{Compute the following: }\\
            $\Delta {\bf p}_a(t) = {\bf p}_{i^{*}}(t) - {\bf p}_a(t-1)$\\ \quad\quad\quad (vector pointing to new anchor position);\\
            
            ${\bf V} = \frac{1}{N}\sum\limits_{i=1}^{N} \overline{\tilde{{\bf V}}}_{i}^w(t)$\\ 
            \quad\quad\quad (vector pointing in mean flow field direction);\\ 
            
            $\beta = \cos^{-1}\left[ \frac{\Delta {\bf p}_a(t) \cdot \left(-{\bf V}\right)}{||\Delta {\bf p}_a(t)|| \times || {\bf V} ||}\right] \times \frac{180^{\circ}}{\pi}$ \\
            \quad\quad\quad (angle formed between the opposite flowfield direction and the direction of the change in anchor positions---see Fig. \ref{fig:Ch5-anchor-angle}); \\

            {\eIf {$0 \leq \beta \leq \beta_{max}$}
            {\textbf{Anchor Node Information Is Updated}\\ 
            $\rho_a(t) = \overline{\tilde{\rho}}_{i^{*}}(t) - 1.98$ (Concentration)\;   
            ${\bf p}_a(t) = {\bf p}_{i^{*}}(t)$ (Position)\; 
            $h_a(t) = h_{i^{*}}(t)$ (Altitude)\; 
            }{
                \textbf{Anchor Node Information Remains Unchanged};\\ 
            }
            
            }
            
            }
            }
            }{
            \textbf{Anchor Node Information Remains Unchanged};\\
            }
            }
            
        }{
        \textbf{Anchor Node Information Remains Unchanged};\\ 
        }}
    \end{algorithm}

    \begin{algorithm}[h]
        \If{\text{Anchor Node Information Remains Unchanged}} 
        {
        $\rho_a(t) = \rho_a(t-1)$ (Anchor Concentration)\; 
        ${\bf p}_a(t) = {\bf p}_a(t-1)$ (Anchor Position)\; 
        }
    \end{algorithm}

    \item State of Obstacles $\left(\{{\bf o}_{u,o}^{i,m} (t) \}\right)$
    
    The last observation vector captures the states of other aerial obstacles in the area: 
    \begin{align}
    {\bf o}_{u,o}^{i,m}(t) = \left[d_{u,o}^{i,m}(t),\psi_{u,o}^{i,m}(t)\right],\;\; \text{for } m \in [1,M]
    \end{align}
    where 
    \begin{itemize}
    \item $d_{u,o}^{i,m}(t)$ is the  relative distance between UAV $i$ and obstacle $m$, and
    \item $\psi_{u,o}^{i,m}(t)$ is the angle between UAV $i$ and obstacle $m$.
    \end{itemize}
    
    The UAVs use this information, along with their proximity sensor readings and collision avoidance protocols, to prevent contact with obstacles. This data is only populated when an obstacle enters the UAV's sensing radius.

\end{enumerate}

\subsubsection{Action Space}
Let $\mathcal{A}$ denote the $N$-dimensional action space of the $N$ UAVs, where $\mathcal{A} = \mathcal{A}_1 \times \mathcal{A}_2 \times \cdots \times \mathcal{A}_N$. The action of UAV $i$ at time-step $t$, denoted as ${\bf a}_{u}^{i}(t) \in \mathcal{A}_i$, is composed of the elements ${\bf a}_{u}^{i}(t) = \left[v_{u}^{i}(t),\omega_{u}^{i}(t) \right]$, which are the adjusted linear and angular velocities of UAV $i$ as discussed in Sec.~\ref{Ch5-UAV-Dynamics}. 

\subsubsection{Reward Space}
Let $\mathcal{R}$ denote the reward space of the UAVs. The goal of each agent is to choose action ${\bf a}_{u}^{i}(t)$ to maximize the cumulative sum of discounted rewards it receives over time. In particular, each agent aims to maximize its accumulated discounted expected return \cite{sutton2018reinforcement} 
\begin{align*}
G_i(t) = \sum\limits_{k=t+1}^{T}\gamma^{k-t-1}R_u^i(k), \;\; i \in [1,N]
\end{align*}
where $\gamma \in [0,1]$ is the discount factor, $T$ is the episode duration, and $R_i(t+1)$ is the immediate reward received by agent $i$ after completing action ${\bf a}_u^i(t)$ during time-step $t$ and transitioning to a new state ${\bf s}_u^i(t+1)$. The goal of the agent is to maximize the ``expected return" (i.e. $\max \mathbb{E}_{\pi_\theta}\left[G_i(t) | S_i(t) = {\bf s}_u^i(t)\right]$, where ${\bf a}_u^i(t) \leftarrow \pi_\theta\left( \cdot | {\bf s}_u^i(t)\right)$ is the action chosen under the parameterized policy $\pi_\theta$. 
Based on the task requirements outlined in Sec.~\ref{sec:Ch5-Tasks}, we design a reward function to evaluate the UAVs' current states and action selections. Traditional approaches to the CPSL problem, such as chemotaxis, anemotaxis, and fluxotaxis, require the agents to maintain a fixed formation while scanning for chemical signatures. However, maintaining this formation in low-concentration areas consumes significant resources to satisfy formation and collision safety constraints. Instead, we propose an adaptive algorithm where the UAVs learn the formation shape only after detecting the plume. Once the plume is detected, the UAVs leverage the formation's structure to their advantage, seeking areas of higher concentration upwind of the plume. The formation's shape is determined by the number of UAVs in the system. For example, if $N = 3$, the UAVs form a triangle; if $N = 4$, the UAVs form a square.

Let $\mathcal{R}$ denote the reward space of the UAVs. The total reward for each UAV $i$ at time-step $t$, denoted as $R_u^{i}(t) \in \mathcal{R}$, is composed of a weighted sum of the following elements: 
\begin{align}
R_u^{i}(t) = \alpha_d r_{d}^{i}(t) &+ \alpha_{\theta}r_{\theta}^{i}(t) + \alpha_{col}r_{col}^{i}(t) + \\ &\alpha_{plume}r_{plume}^{i}(t) + \alpha_{upwind}r_{upwind}^{i}(t) \nonumber 
\end{align}
for $i \in [1,...,N]$, and $\{\alpha_d,\alpha_\theta,\alpha_{col},\alpha_{plume},\alpha_{upwind}\}$ are the reward weights. The reward function must account for two primary objectives: formation control and CPSL guidance. First, we will address the rewards designed to guide the UAVs in establishing and maintaining an optimal formation. Next, we will discuss the rewards that leverage sensor data to direct the UAVs toward the emitter.

\noindent \textbf{Formation Control Rewards}

The UAV formation emerges implicitly as a result of the following reward structure. Specifically, the distance, angle, and collision rewards are designed to encourage UAVs to stay close to the anchor point while maintaining safe spacing to reduce collision risk. Consequently, as demonstrated in the animation in Sec.~VI, the UAVs self-organize into a roughly circular formation and exhibit coordinated movement---rotating around the anchor while advancing toward the emitter. This dynamic formation provides both flexibility and robustness, enabling the UAVs to cope with wind turbulence and transient sensor readings within the plume.

\begin{enumerate}[label=\arabic*.,leftmargin=*,listparindent=1.5em] 
    \item Distance Reward $\left(r_{d}^{i}(t) \right)$


    The distance reward encourages UAVs to maintain a flexible yet cohesive formation. Instead of a fixed ideal distance, a range defined by $d_{ideal\_min}$ and $d_{ideal\_max}$ allows for adaptive spacing around the centroid. UAVs within this range receive a small positive reward, while those outside are penalized proportionally to their deviation. The reward is defined as:
    \begin{align}
    r_d^{i}(t) = 
    \begin{cases}
    r_{\text{in}}, & \text{if } d_{ideal\_min} \leq d_{u,c}^{i}(t) \leq d_{ideal\_max} \\
    -\left(k_{d1} \cdot \delta_d^{i}(t) \right), & \text{otherwise}
    \end{cases}
    \end{align}
    where 
    \begin{itemize}
        \item $r_{\text{in}}$ is a small positive reward for staying within range,
        \item $k_{d1}$ is a penalty coefficient,
        \item $\delta_d^{i}(t) = \min\left( |d_{u,c}^{i}(t) - d_{ideal\_min}|, |d_{u,c}^{i}(t) - d_{ideal\_max}| \right)$ is the deviation from the closest boundary,
    \end{itemize}
    
    This formulation preserves formation structure while granting UAVs autonomy to adjust spacing based on the search context. By operating within a defined range, the UAVs can dynamically expand or contract, thereby modulating the scope and density of the search as needed.

    \item Interior Angles of the Formation Reward $\left(r_{\theta}^{i}(t)\right)$
    
   To promote a stable encirclement of the emitter source, the interior angle, $\gamma_{i,j}$ formed between the line connecting UAV $i$ to the centroid and the line connecting a neighboring UAV $j$ to the centroid should ideally be $\frac{2\pi}{N}$ (see Fig. \ref{fig:Ch5-RelativeDefinitions}(b)). Thus, each UAV $i$ must maintain an interior phase difference of $\frac{2\pi}{N}$ relative to its adjacent neighbors, UAV $j$ and UAV $k$. The reward function is defined as: 
    \begin{align}
    r_{\theta}(t) &= k_{\theta_1}r_{\theta_1}(t) + k_{\theta_2}r_{\theta_2}(t)
    \end{align}
    where $k_{\theta_1}$ and $k_{\theta_2}$ are penalty coefficients, and 
    
    \begin{itemize}
    \item $r_{\theta_1}(t) = \exp\left(- \biggr |\gamma_{i,j}(t) - \frac{2\pi}{N} \biggr | \right) + \exp\left(-\biggr |\gamma_{i,k}(t) - \frac{2\pi}{N} \biggr|  \right) - 2 $,
    \item 
    $r_{\theta_2}(t) = \exp\left(-|\gamma_{i,j}(t) - \gamma_{j,k}(t)| \right) - 1 $.
    \end{itemize}

     \item Collision Reward $\left(r_{col}^{i}(t) \right)$
    
    A penalty is applied to any UAV that collides with another UAV or an aerial obstacle, and is defined as: 
    \begin{align}
    r_{col}^{i}(t) = \sum\limits_{\substack{i' = 1 \\ i'\ne i}}^N r_{col}^{i,i'}(t) + \sum\limits_{m = 1}^{M} r_{col}^{i,m}(t)
    \end{align}
    where 
    \begin{alignat*}{2}
    r_{col}^{i,i'} &= -k_{col,UAV},\;\;\;\; &&\text{if } ||{\bf p}_{u}^{i}(t) - {\bf p}_{u}^{i'}(t)|| \leq 2r_{u} + \varepsilon \\
    r_{col}^{i,m} &= -k_{col,Obs},\;\;\; &&\text{if } ||{\bf p}_u^i(t) - {\bf p}_o^m(t)|| \leq r_{u} + r_{o} + \varepsilon 
    \end{alignat*}
    for $i' \in [1,N]{\backslash}i$ and $m \in [1,M]$. The first condition penalizes UAVs that collide with each other, while the second condition penalizes UAVs that collide with obstacles. 
    \end{enumerate}
    \noindent \textbf{CPSL Guidance Rewards}
    
    This reward structure encourages UAVs to maintain proximity to the anchor and to assist in updating its position upwind, thereby guiding the formation toward the emitter. The anchor update rule also prevents overshooting in the upwind direction, helping the anchor remain within the plume. As a result, there is no need to include a terminal task reward to signify successful localization. In fact, we found that such a reward could be counterproductive, as it inadvertently encouraged UAVs to ``memorize'' the emitter's location instead of relying on real-time sensor data.
    
    \begin{enumerate}[label=\arabic*.,leftmargin=*,listparindent=1.5em] 
    \item Plume Trace Reward $\left(r_{plume}^{i}(t) \right)$

    First, the UAVs must quickly identify the plume. This is achieved by rapidly establishing anchor node information. Once an anchor is identified, the following reward incentivizes the UAVs to move toward the anchor as swiftly as possible. While traversing, they continue collecting sensor data, and it is likely that a new anchor may be established by one of the UAVs, prompting a change in direction to pursue this new anchor. This reward structure encourages the UAVs to stay close to the established anchor, reducing the likelihood of drifting away from the plume. This behavior is further reinforced by the formation rewards. 
    
    \begin{align*}
    r_{plume} = \begin{cases}
    -\eta d_{a}^{i}(t) & \text{if $\rho_a(t) \neq 0$} \\ 
    -\eta d_{max} & \text{otherwise}
    \end{cases}
    \end{align*}
    where $d_{max}$ is the maximum separation distance between two UAVs within the simulation environment. 

    \item Upwind Reward $\left(r_{upwind}^{i}(t) \right)$

    In addition to encouraging the UAVs to converge toward an established anchor, it is important to incentivize them to continue moving upwind of the plume. Without this incentive, UAVs may linger near the anchor to minimize negative rewards. To address this, a small reward, on the same scale as $r_{plume}$, is given when the UAVs move in the upwind direction of the plume. Whether a UAV is moving upwind is determined using the same method described in Algorithm \ref{anchor-update}. The pseudocode for this update is provided in Algorithm \ref{code:R_upwind}. 

    \begin{algorithm}
    \caption{Update Rule for $r_{upwind}^i(t)$}
    \label{code:R_upwind}
    {\text{Compute the following: }\\
    $\Delta {\bf p}_{i,a}(t) = {\bf p}_{i}(t) - {\bf p}_a(t)$\\ \quad\quad\quad (vector pointing from the current anchor node to agent $i$'s current position);\\
            
    ${\bf V} = \frac{1}{N}\sum\limits_{i=1}^{N} \overline{\tilde{{\bf V}}}_{i}^w(t)$\\ 
            \quad\quad\quad (vector pointing in mean flow field direction);\\ 
            
    $\beta_i = \cos^{-1}\left[ \frac{\Delta {\bf p}_i(t) \cdot \left(-{\bf V}\right)}{||\Delta {\bf p}_i(t)|| \times || {\bf V} ||}\right] \times \frac{180^{\circ}}{\pi}$ \\
            \quad\quad\quad (angle formed between the opposite flow-field direction and the direction of $\Delta {\bf p}_{i,a}(t)$); \\

    {\If {$\left[0 \leq \beta_i \leq \beta_{max} \textbf{\quad AND \quad } \rho_a(t) \neq 0 \right]$ }
            {$r_{upwind}(t) = \frac{r_{plume}}{4}$}
    {\ElseIf{$\left[\beta_{i} > \beta_{max} \textbf{\quad AND \quad } \rho_a(t) \neq 0 \right]$}
            {$r_{upwind}(t) = \frac{r_{plume}}{2}$}
    
    {\Else{$r_{upwind}(t) = r_{plume}$}
    }
    }
    }
    }
    \end{algorithm}

\end{enumerate}

\subsubsection{Termination Conditions}
\label{sec:termination}
Thanks to the anchor node design, an explicit termination condition is no longer required. As shown in the animation referenced in Sec.~\ref{sec:Ch5-NumericalResults}, once the anchor node reaches the boundary of the plume at the designated altitude, it naturally stops progressing upwind. This occurs because the UAVs are unable to identify any upwind location with a concentration exceeding the predefined threshold $\rho_{\mathrm{th}}$. To further improve the efficiency of the CPSL process, we introduce a dynamic stopping criterion based on the movement of the centroid. Specifically, if the centroid’s position remains relatively stable within a given time window, the system can terminate the tracing phase and declare the emitter’s estimated location, without waiting for the time limit to be reached.

By examining the plume from the left view (i.e., the middle plot of the plume example\footnote{Animation of this plume example is posted on YouTube: \url{https://www.youtube.com/watch?v=_qtImvQBi4w}}), we observe that the plume boundary at certain altitudes (e.g., 5 m) does not perfectly align with the true emitter location along the X-axis. Even with accurate tracing, the anchor eventually stabilizes near the plume boundary but exhibits a small offset from the actual emitter position. This offset tends to increase at higher altitudes and under stronger wind conditions. To improve localization accuracy, we introduce a corrective offset in the upwind direction, applied after the emitter location is initially identified. In this study, the offset is determined heuristically based on a fixed average wind speed and search altitude. For other environmental configurations, this offset could be estimated using a regression-based approach.

\subsection{Deep Reinforcement Learning (DRL) Architecture}
The DRL architecture follows a similar design to our previous work \cite{sherman2024counter}. It consists of three main components: (1) a DeepSet neural network architecture \cite{pierre2023multi} for pre-processing the time-varying observation vectors of the UAVs, (2) a proximal policy optimization (PPO) algorithm, and (3) a collision safety protocol that overrides any PPO-derived actions that would result in a collision between UAVs or between UAVs and obstacles. Through repeated interactions with the environment, the UAVs learn an optimal policy that enhances the emitter localization rate while adhering to collision and formation constraints. 

In comparison to our previous work \cite{sherman2024counter}, the DeepSet architecture presented here is composed of six fully connected neural networks that separately process the variable-length observation sets $\{{\bf o}_{u}^{i,i'} (t) \}, \{ {\bf o}_{u,s}^{i,i'} (t)\},$ and $\{{\bf o}_{u,o}^{i,m}(t) \}$. Each observation type is passed through a shared multilayer perceptron (MLP) block consisting of two linear layers with ReLU activation and 256 neurons, followed by a permutation-invariant pooling operation, and a second MLP block of similar structure to produce fixed-size embeddings. These embeddings are then concatenated with raw agent-specific observations—such as the agent state ${\bf s}_{u}^{i}$, local sensor measurements ${\bf o}_{u,s}^{i}$, and the anchor observation ${\bf o}_{a}^{i}$—to form a unified feature vector. This vector is subsequently processed by a two-layer fully connected network (Linear - ReLU - Linear, 256 neurons), yielding shared representations that are used for both the policy and value function outputs. This design choice is motivated by the simulation setup, which assumes a single emission source and allows at most one anchor node per timestep, which ensures that these observations remain consistent in structure and do not require set-based processing.

\section{Simulation and Performance Evaluation}
\label{sec:Ch5-NumericalResults}

To reduce computation time, the plume model is generated prior to initiating the MARL training. The system consists of $N = 3$ UAV agents and $M = 5$ aerial obstacles. Additional environment and algorithm parameters are listed in Table~\ref{table:Ch5-EnvParameterTable}. The agents are initially deployed to perform a vertical sweep across the environment grid, perpendicular to the plume's centerline. The UAVs traverse at a fixed velocity until they either detect the plume or reach the environment boundary at which point they reverse direction and continue exploration. The agents are tasked with localizing the emitter within a $200 \times 200$ m$^2$ grid over a simulation duration of $T = 160$ seconds (or 3200 time steps). 

\begin{table}[ht]
\centering
\caption{Simulation Parameter Values}
\label{table:Ch5-EnvParameterTable}
\begin{tabular}{|l|c|}
\hline
\textbf{Description} & \textbf{Value}\\
\hline
Geographical Area ($X \times Y$) & $200\; \text{m} \times 200\; \text{m}$ \\
 \hline 
Grid element size ($\Delta x \times \Delta y$) & $2\; \text{m} \times 2\; \text{m}$ \\
\hline
Number of UAVs ($N$) & $3$ \\
\hline
Number of Obstacles ($M$) & $5$ \\
\hline
UAV Object Sensing Radius ($r_{s}$)  & $20$ m \\
\hline 
Collision Safety Distance ($\epsilon$)  & 1 m \\
\hline 
Declaration Boundary Radius & 4 m \\
\hline 
Maximum Length of Episode ($T$) & 160 sec. \\ 
\hline 
Time-step size ($t_s$) & 0.05 sec. \\
\hline 
Replay Buffer Size & 16384 \\ 
\hline 
Mini-batch Size & 1024 \\ 
\hline
Learning Rate for Actor \& Critic ($\vartheta$) & $8.4856 \times 10^{-5}$\\
\hline
Concentration Threshold ($\rho_{\mathrm{th}}$) & 0.52 ppm\\
\hline 
\end{tabular}
\end{table}

The training process was conducted on a machine equipped with an NVIDIA RTX 5090 GPU, 32 GB of RAM, and running Ubuntu 24.04.3 LTS. The reinforcement learning experiment employed the PPO algorithm implemented through the Ray RLlib framework, with a custom model architecture based on DeepSet. The training session was executed locally and concluded after 250 iterations. Over a total training time of approximately two hours, the agent interacted with the environment for 4.8 million environment steps, corresponding to approximately 14.4 million agent steps. The entire training process used the same environment configuration described in Table~\ref{table:Ch5-PlumeParameterTable}.

To evaluate the adaptability of the trained model to different environments, we train it using a fixed emitter location and wind condition (i.e., ${\bf p}_{\mathrm{es}} = (60, 120)$ and $[a, b, G] = [0.005, 0.02, 1]$). The model is then tested under various emitter locations and wind conditions, as listed in Table~\ref{table:TestCh5-PlumeParameterTable}, to assess its generalization capability.

\begin{table}[h!]
\centering
\caption{Test Plume Parameter Values}
\label{table:TestCh5-PlumeParameterTable}
\begin{tabular}{|c|c|}
\hline
\textbf{Notation} & \textbf{Value}\\
\hline 
Emitter Location & 
(80, 60) m \\
& (60, 120) m \\
\hline 
\multirow{3}{*}{Noise Constants $\left([a,b,G]\right)$} & $[0.005,\; 0.02,\; 1]$ \\
& $[0.005,\; 0.02,\; 3]$ \\
& $[0.005,\; 0.02,\; 5]$ \\
\hline 
\end{tabular}
\end{table}



\begin{figure}[ht]
\centering
    \includegraphics[width=0.4\textwidth]{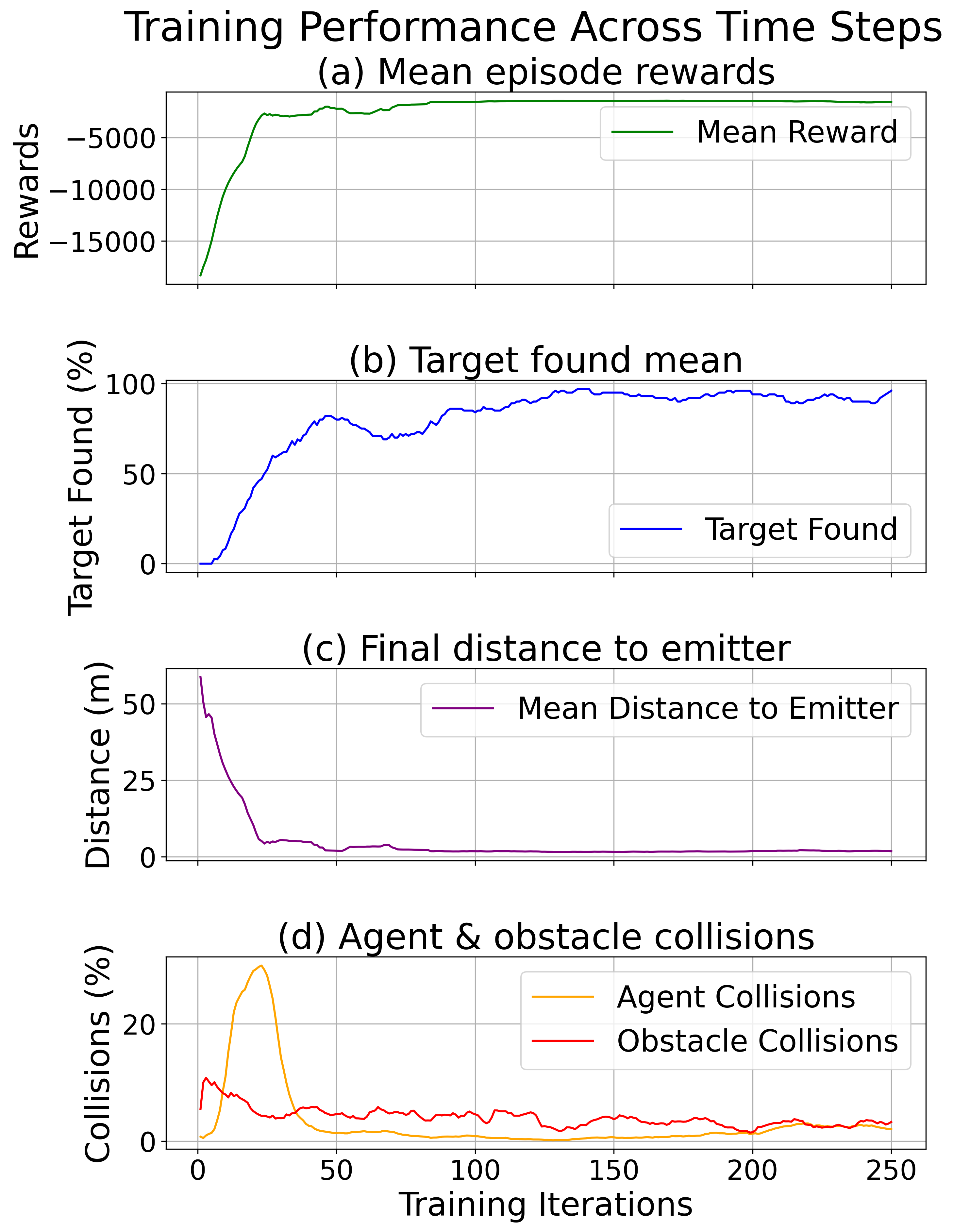}
    \caption{Simulation results for Plume Example 1: (a) Mean Total Reward $R_u^{i}$ received per agent, (b) Mean Target Found Rate, (c) Mean Final Centroid Error, (d) Mean Agent-to-Agent and Agent-to-Obstacle Collision Rate.}
    \label{fig:Ch5-Results_pt1}
\end{figure}

\begin{figure}[ht]
\centering
    \includegraphics[width=0.48\textwidth]{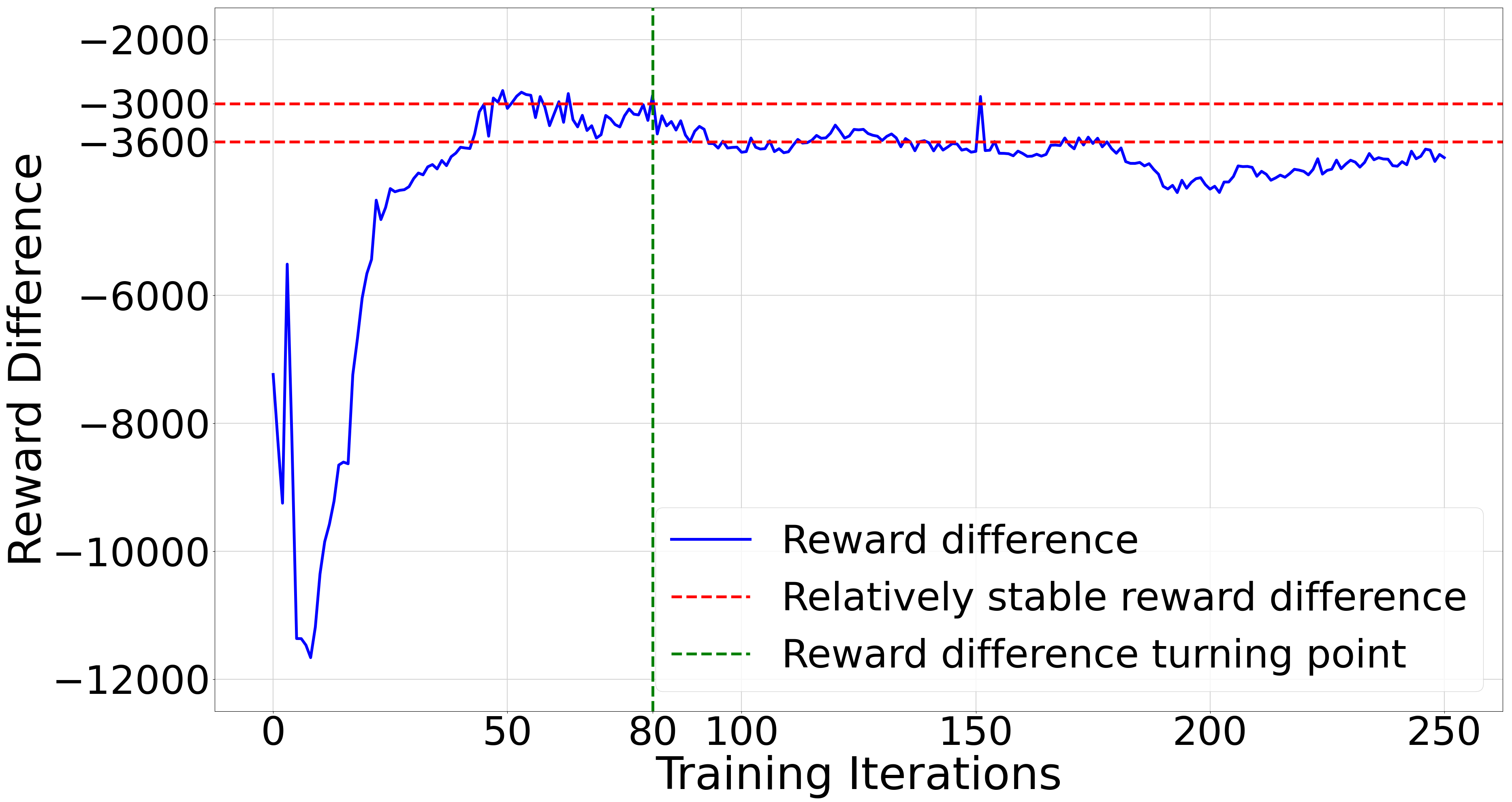}
    \caption{Reward difference during training.}
    \label{fig:Ch5-Results_pt2}
\end{figure}
\subsection{Training Performance}
\subsubsection{Mean Total Reward Received per Agent}
This metric measures the total reward $R_{u}^{i}$ per agent during each episode, averaged over a training iteration. Fig. \ref{fig:Ch5-Results_pt1}(a) shows the total reward received per agent. The reward function, which includes multiple components influencing each other, was found to be learned sequentially by the UAVs over several simulation runs. Initially, the UAVs quickly learned how to configure themselves into formation, followed by determining optimal actions when an anchor was established, and finally, how to best move the formation upwind toward the emitter. Observing the simulation animation \footnote{Animation of CPSL phases is posted on YouTube: \url{https://www.youtube.com/watch?v=98LmWbcowOk}}, we see that the UAVs first converge toward the most recent anchor, configure themselves into formation, and then make gradual progress upwind. 

\subsubsection{Mean Success Rate for Emitter Localization}
This metric is calculated as the running average of the total number of episodes where the emitter was successfully localized (i.e., ${\bf p}_{es} \in \mathcal{D}$), divided by the total number of episodes completed up to each training iteration. Here, $\mathcal{D}$ denotes a circular region with a radius of 5 meters centered at the emitter source. The converged results in Fig. \ref{fig:Ch5-Results_pt1}(b) demonstrate that the UAVs successfully declared the emitter to be within the final declaration region $\mathcal{D}$ in approximately 95\% of the episodes completed.   

\subsubsection{Mean Final Centroid Error}
This metric measures the average Euclidean distance between the centroid of the UAV formation and the true emitter source location, expressed as $|| {\bf p}_{es} - {\bf p}_{c}(T_{end})||$, where ${\bf p}_{c}(T_{end})$ is the final declared location based on the centroid. Fig. \ref{fig:Ch5-Results_pt1}(c) plots the average Euclidean distance error between the final centroid location and the true emitter location. The results indicate that, on average, the UAVs completely surrounded the emitter source by the end of the simulation. The centroid of the formation is within 5 meters of the emitter at the time of episode termination, which is less than the maximum dimensions of the final formation. This value depends on several factors, including the formation constraints, wind speed, detection thresholds, and the transient behavior of the plume. In practice, when the declared location is within a few meters of the actual emitter source, additional handheld instruments—such as infrared camera \cite{CDPHE_AIMM}—can be used to confirm the precise emitter location.

\subsubsection{Mean Number of Collisions between Agents and Obstacles} This metric measures the average number of collisions between UAVs and obstacles across episodes during each training iteration. Fig.~\ref{fig:Ch5-Results_pt1}(d) illustrates the ratio of episodes involving collisions to the total number of episodes per iteration. Both the quantitative results and the simulation animations indicate that the UAVs effectively learned to avoid obstacles. When an obstacle approaches the formation, the UAVs adapt by temporarily breaking formation to avoid a collision and subsequently returning to formation once the threat has passed. Future work will focus on further reducing the collision rate toward zero—for example, by integrating a Quadratic Programming (QP) solver—and refining the formation control reward functions to better balance exploration capability with collision avoidance.
\subsection{Validation for Model Selection} \label{model_selection}

In practical applications, a specific checkpoint of the trained model is typically selected for deployment. However, overfitting often occurs during the deep learning training process, making it important to identify the most appropriate checkpoint for deployment. The model selection strategy used in this work is a variation of the classical early stopping technique \cite{Caruana2000OverfittingIN}, which monitors validation performance to prevent overfitting.

Specifically, we generate a new scenario, \textit{small\_100\_100}, as the validation environment during training (i.e., ${\bf p}_{\mathrm{es}}=(100,100)$ and $[a, b, G] = [0.005, 0.02, 3]$), whose data distribution differs from that of the training scenario. Every five episodes, a validation process is performed to evaluate the model by computing the reward on the validation scenario, denoted as $R_{valid}$. We also compute the reward on the training environment, denoted as $R_{train}$, and calculate the difference between the training and validation rewards, $R_{diff}$, as defined in Equation~\eqref{eq_R_diff}.

\begin{equation}
    R_{diff} = R_{train} - R_{valid}
    \label{eq_R_diff}
\end{equation}

At the early stage of training, as the model gradually converges, the absolute value of $R_{diff}$ decreases. However, as training continues, the model begins to overfit the training environment, causing the reward gap between the training and validation environments to increase. Consequently, the absolute value of $R_{diff}$ starts to grow. We define the point where the absolute value of $R_{diff}$ begins to increase as the \textit{turning point}. The model checkpoint corresponding to this turning point is considered suitable for deployment.

Following this approach, we use scenario \textit{no\_60\_120} as the training environment (i.e., ${\bf p}_{\mathrm{es}}=(60,120)$ and $[a, b, G] = [0.005, 0.02, 3]$) and select the checkpoint at episode 80 for deployment. The curve of $R_{diff}$ is shown in Figure~\ref{fig:Ch5-Results_pt2}.

\subsection{Test Performance}
\label{sec:test_performance}
\begin{figure}[ht]
   \centering
    \includegraphics[width=0.48\textwidth]{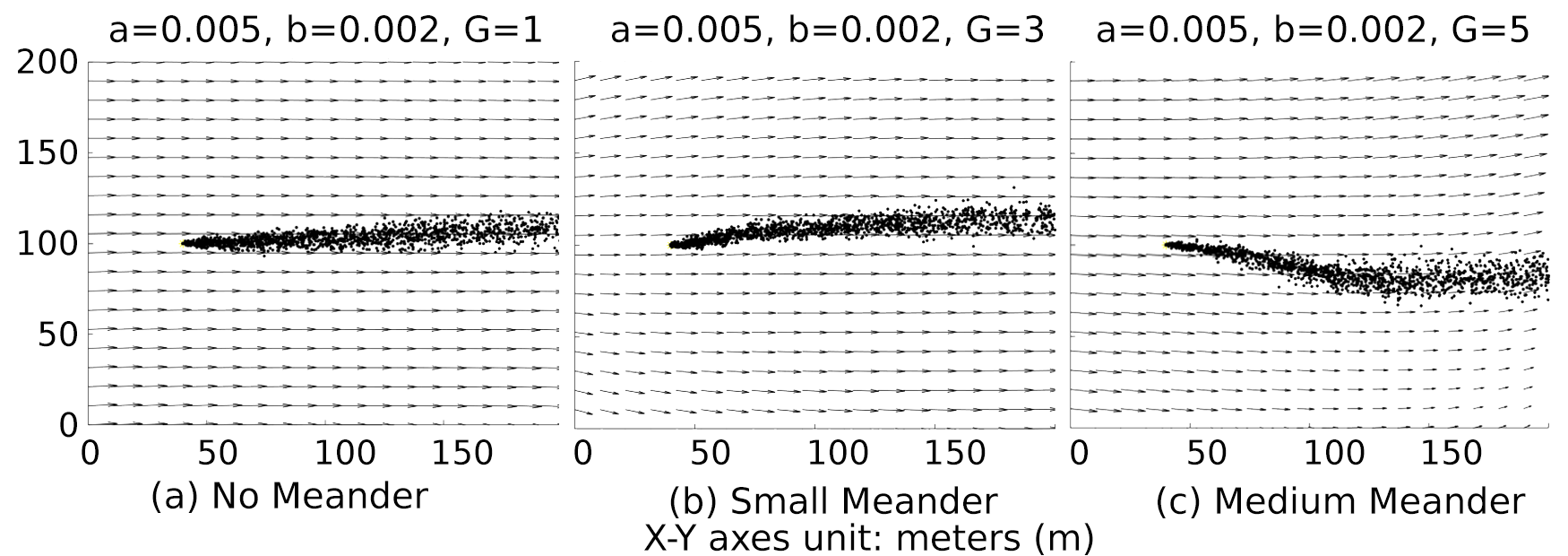}
    \caption{Plume distribution with different wind conditions}
    \label{fig:plume_diagram}
\end{figure}

To evaluate the robustness and generalization capability of our method, we construct six distinct testing datasets by varying both wind conditions and emitter locations. Specifically, we define three wind scenarios—no meander, small meander, and medium meander—which simulate increasing levels of wind turbulence and directional variability (see Fig.~\ref{fig:plume_diagram}). This figure illustrates plume dispersion under three wind scenarios: (a) no meander with steady wind, (b) small meander with slight wind oscillations, and (c) medium meander with stronger fluctuations causing wider dispersion. These wind profiles impact the dispersion of the plume and consequently influence the agent's search trajectory. For each wind condition, we generate two test datasets with different emitter positions. This results in a total of six unique test cases. All other plume simulation settings remain the same as those specified in Table~\ref{table:Ch5-PlumeParameterTable}, except for the Emitter Source Location and the Colored Noise Constants, which are varied as shown in Table~\ref{table:TestCh5-PlumeParameterTable}.

To further introduce variability and evaluate performance under diverse environmental realizations, each test episode is initialized by choosing random starting positions for the UAV group. In addition, the positions of all aerial obstacles are also randomly generated for each episode. We conduct 100 independent test episodes for each of the six scenarios, capturing a wide range of conditions and agent behaviors. This comprehensive testing protocol enables us to assess the agents' ability to adapt not only to dynamic environmental conditions but also to spatial changes in the target location.  

According to the method described in Section~\ref{model_selection}, we observed the reward difference curve and identified episode 80 as the turning point (marked by the green dashed line in Fig.~\ref{fig:Ch5-Results_pt2}). Therefore, the model at this checkpoint was selected as the test model. It should be noted that, in the simulation, we evaluated several checkpoints near the turning point (both before and after episode 80). Based on comparisons across different combinations of training and validation scenarios, the performance of the selected checkpoint is within $90\%$ of the optimal performance and, in some cases, significantly outperforms models that exhibit overfitting. The test results of the checkpoint-80 model across six scenarios are summarized in Table~\ref{tab:test_summary}. The offset is introduced in Sec.~\ref{sec:termination}.

\begin{table}[h]
\centering
\caption{Success rate and location error (i.e, final distance to emitter) under various environment conditions}
\label{tab:test_summary}
{\small
\begin{tabular}{|c|c|c|c|c|c|c|}
\hline
\multicolumn{2}{|c|}{\multirow{2}{*}{\textbf{Scenario Setup}}} & 
\multirow{2}{*}{\makecell[c]{\\[0.5ex] \textbf{Succ.} \\ \textbf{Rate}}} & 
\multicolumn{4}{c|}{\textbf{Mean Location Error (m)}} \\
\cline{4-7}
\multicolumn{2}{|c|}{} & & \multicolumn{2}{c|}{\textbf{No Offset}} & \multicolumn{2}{c|}{\textbf{With Offset}} \\
\cline{1-2} \cline{4-7}
\makecell[c]{\textbf{Wind} \\ \textbf{Meander}} & 
\makecell[c]{\textbf{Emitter} \\ \textbf{Location}} & 
& \textbf{All} & \textbf{Succ.} & \textbf{All} & \textbf{Succ.} \\
\hline
\multirow{2}{*}{No}     & [80, 60]        & 1.00 & 2.34 & 2.34 & 1.55 & 1.55 \\
                        & [60, 120]       & 1.00 & 2.17 & 2.17 & 1.52 & 1.52 \\
\hline
\multirow{2}{*}{Small}  & [80, 60]        & 1.00 & 2.16 & 2.16 & 1.76 & 1.76 \\
                        & [60, 120]       & 0.96 & 2.61 & 2.40 & 1.7 & 1.48 \\
\hline
\multirow{2}{*}{Medium} & [80, 60]        & 0.99 & 2.26 & 2.24 & 1.60 & 1.58 \\
                        & [60, 120]       & 0.73 & 11.10 & 2.27 & 10.25 & 1.65 \\
\hline
\end{tabular}
}
\end{table}

Table~\ref{tab:test_summary} summarizes the testing success rates and mean localization errors (i.e., final distances to the emitter) across six test configurations that vary by wind conditions and emitter placement. Overall, the algorithm demonstrates strong performance in stable wind environments. Under no-meander conditions, success rates remain consistently high ($>0.99$), and mean localization errors are typically below 2.4 m. However, as wind meander increases, the success rate degrades, particularly for the emitter location at $[60,120]$. In the medium-meander scenario, success rates drop to as low as 0.73, and the mean localization errors increase noticeably, highlighting the greater difficulty of emitter localization under turbulent plume dispersion.

Despite these challenges, the algorithm still exhibits a reasonable degree of robustness. Under both small and medium meander conditions, it often converges near the emitter during successful episodes, as reflected in the ``Succ.'' columns, where the mean localization errors remain relatively low (e.g., 1.65 m under medium meander with offset). A comparison between the ``All'' and ``Succ.'' columns further indicates that most localization errors arise from unsuccessful episodes. Nevertheless, even in these cases, the declared emitter location typically provides a reasonably close estimate. When combined with an infrared camera, this estimate can still effectively support emitter source identification in practical applications.


\subsection{Behavior Analysis}
Through animation visualization\footnote{Animation of CPSL phases is posted on YouTube: \url{https://www.youtube.com/watch?v=98LmWbcowOk}}, we are able to test the trained network and observe the three phases of the CPSL problem: \textit{seek}, \textit{trace}, and \textit{localize}.

\begin{figure}[h!]
   \centering
    \includegraphics[width=0.49\textwidth]{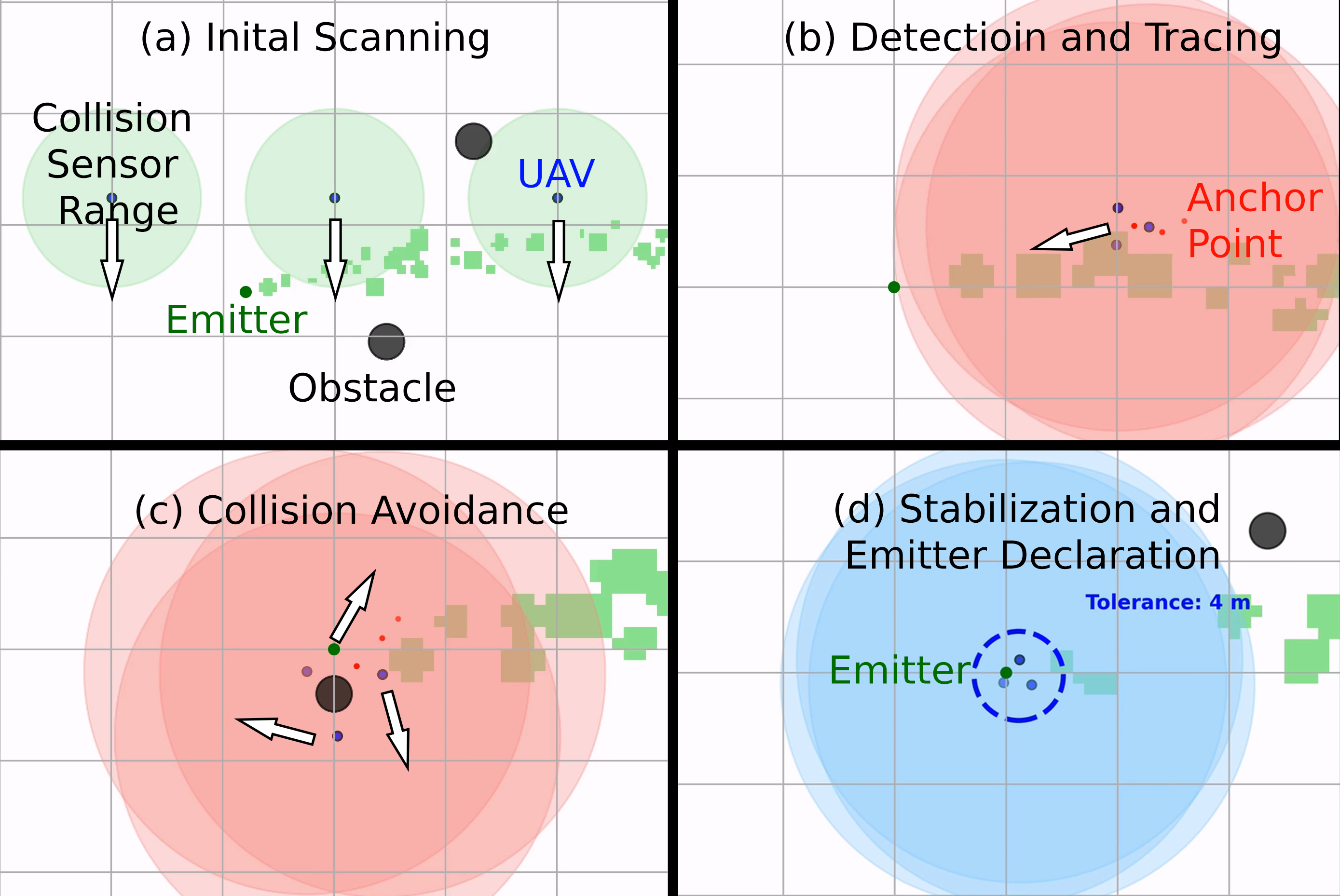}
    \caption{Example of different UAVs behaviors during a successful episode}
    \label{fig:Ch5-SwarmBehavior}
\end{figure}

Fig.~\ref{fig:Ch5-SwarmBehavior} illustrates the UAVs' behavior as the system progresses through each phase of the CPSL task. In the initial phase, subplot(a), all UAVs begin from a fixed starting position and advance forward until they either detect the chemical plume or reach the boundary of the environment. If no detection occurs, the UAVs reverse direction and adjust their headings to continue the search (see 5:46 in the animation). In trace phase, subplot(b), once at least one UAV detects the plume, the system transitions to plume tracing by establishing an initial anchor node. The UAVs then form a coordinated formation around this anchor, which takes a triangular shape in this case due to the use of three UAVs. They continuously update the anchor position in the upwind direction based on concentration measurements. In subplot(c), when a moving obstacle enters the observation range of a UAV, the affected UAV either increases its distance from the group centroid or maneuvers around the obstacle using the shortest and fastest path available. Once the obstacle is avoided, the UAV promptly resumes coordination with the group to re-establish the intended formation abround the anchor point (see 3:10 in the animation). In the final phase, subplot(d), the centroid of the UAV formation stabilizes, and the algorithm declares the estimated emitter location. If this declared location lies within the predefined region $\mathcal{D}$, the episode is considered successful.

The animation, which presents episodes across two emitter locations and three wind conditions, reveals consistent UAV behavior. Once the UAVs establish the desired formation, they begin rotating in a clockwise direction while maintaining the formation and tracing the plume. This emergent behavior is a direct result of the reward function design. In order for a UAV's measurement to trigger an update of the anchor node, the UAV must position itself upwind of the plume while satisfying the prescribed distance and angular constraints of the formation. This requirement leads to a clockwise rotational pattern, as each UAV competes to advance upwind and assume the ``leader'' position. The resulting motion incrementally drives the entire formation closer to the plume source.

\subsection{Comparison with fluxotaxis Method}
To compare the performance of our proposed CPSL method against the traditional fluxotaxis approach, we design a series of experiments under three wind conditions, discussed in Sec.~\ref{sec:test_performance}. In each scenario, both methods are initialized around the same UAV starting position [130, 60] to ensure a fair comparison. In particular, for the MARL method, agents are required to detect a sufficient signal threshold before transitioning to the next control phase; otherwise, they remain in place. The fluxotaxis agents are initialized at the starting location using the Lennard-Jones-based formation control (with parameters following Sec.~9.4.1 in~\cite{spears2012physicomimetics}) and utilize double integrator dynamics. The fluxotaxis control vector is computed as the dot product between the current agent and the neighboring agents position with the product of the neighboring agents concentration and wind speed vector. The direction of the fluxotaxis vector is selected based on Algorithm in Listing 9.9 in~\cite{spears2012physicomimetics}.
The control vector---based on the summation of the formation control vector (gain set to 10) and fluxotaxis control vector (gain set to 1)---is designed such that the agents keep formation but will still trace the plume effectively. Additionally, an artificial drag term is introduced to help deal with the noisy acceleration vector.

\begin{figure*}[t]
    \centering
    \includegraphics[width=\textwidth]{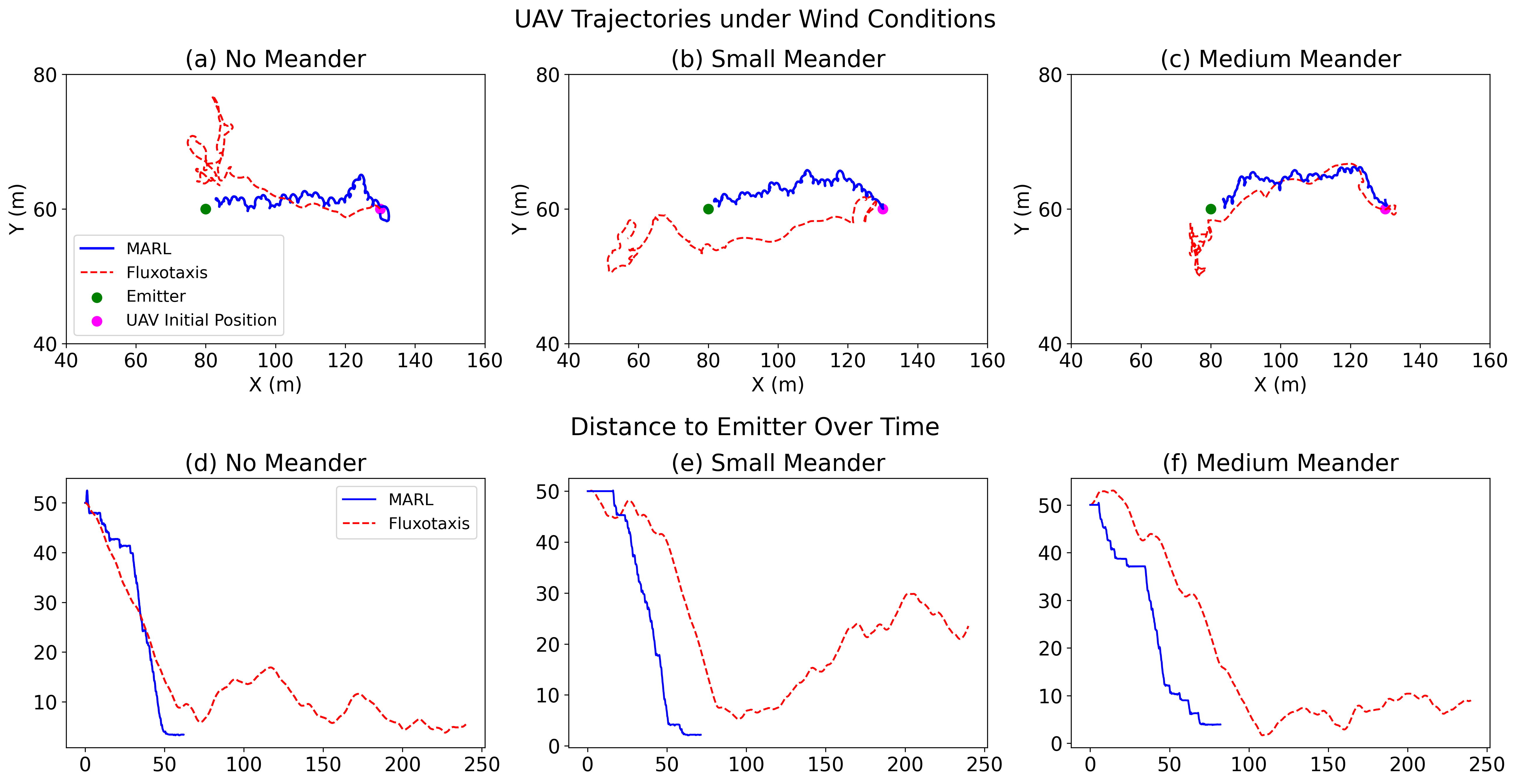}
    \caption{Trajectory and location error evolution for MARL and fluxotaxis under varying wind meandering conditions}
    \label{fig:test_traj_error}
\end{figure*}

\begin{figure}[h]
    \centering
    \includegraphics[width=0.48\textwidth]{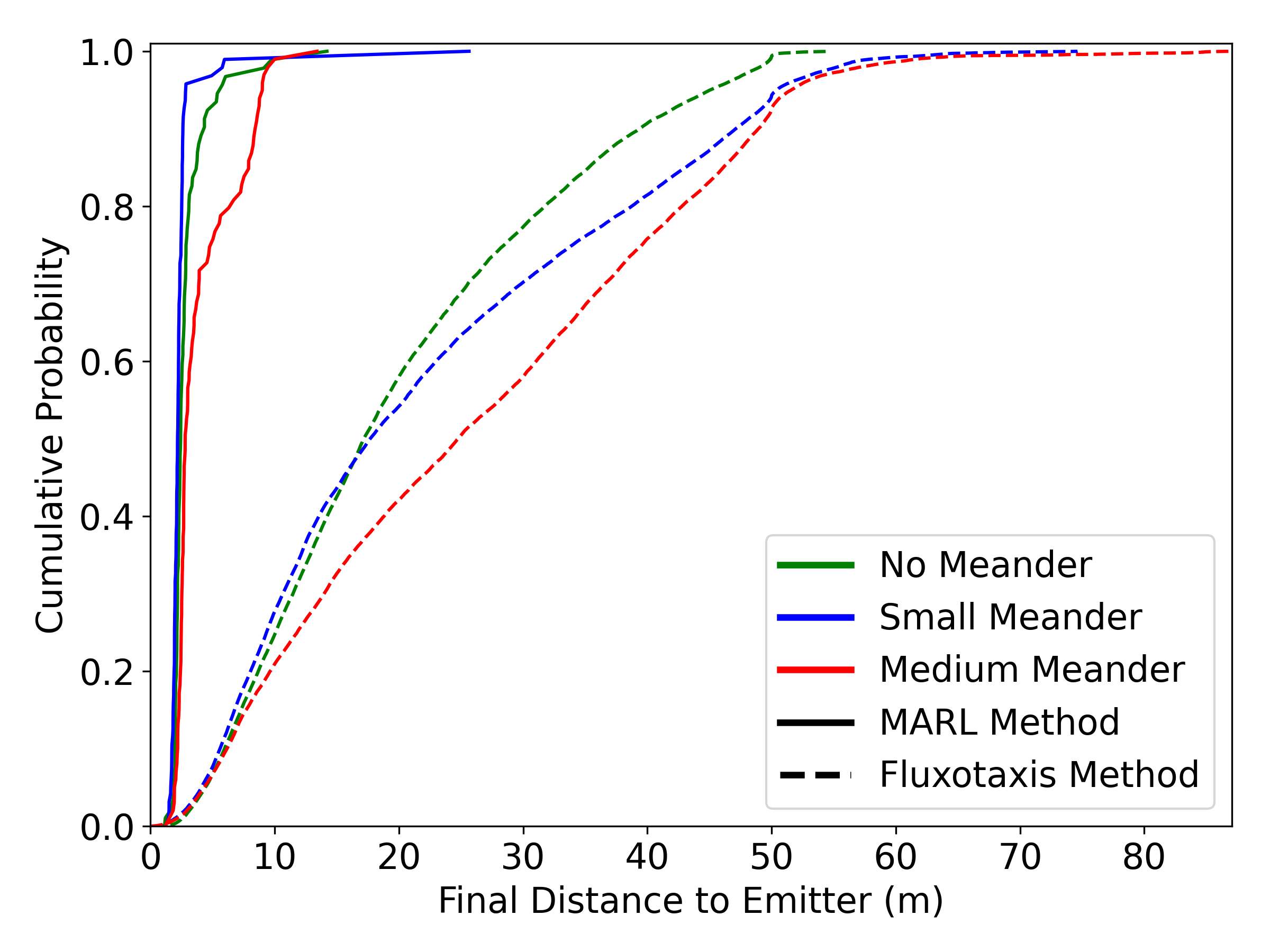}
    \caption{CDF of final distance to emitter for MARL and fluxotaxis (100 tests per case)}
    \label{fig:test_cdf}
\end{figure}

{Fig.~\ref{fig:test_traj_error} shows the spatial trajectories and distance profiles of a representative centroid under three wind conditions for both the MARL and fluxotaxis methods. Subplots (a)–(c) illustrate the spatial paths, where the fluxotaxis trajectories (solid red lines) exhibit substantial lateral deviations and erratic movement, particularly as wind meandering increases. In contrast, the MARL trajectories (solid blue lines) are smoother and more directed toward the emitter. Subplots (d)–(f) display the time evolution of the centroid's distance to the emitter. The MARL agents consistently maintain tighter approach paths and show greater adaptability to wind disturbances. Notably, the MARL trajectories exhibit an initial plateau phase corresponding to a stationary period, during which agents await sufficient plume concentration to initiate movement. After this phase, the MARL-controlled agents rapidly converge on the target location.

Fig.~\ref{fig:test_cdf} presents the cumulative distribution function (CDF) of the final distances to the emitter for both methods under varying wind conditions. Across all scenarios, the MARL method achieves a steeper and more left-shifted CDF, indicating a greater proportion of trials with shorter final distances. This highlights the superior precision and consistency of the MARL approach in target localization. Although both methods exhibit similar overall trends in approaching the emitter, the MARL approach consistently demonstrates higher efficiency and accuracy across all tested conditions.

To ensure full reproducibility, we have made all our codes for MARL and fluxotaxis methods and a detailed README file available in a public GitHub repository \footnote{https://github.com/Shao-wireless-lab/CPSL-Sim}.

\section{Conclusion and Future Work}
\label{sec:Ch5-Conclusion}
In this work, we developed a reinforcement learning framework based on a CTDE MARL architecture to address the key challenges of cooperative sensing in solving the CPSL problem. Simulation results demonstrate the framework’s ability to effectively manage practical challenges such as wind dynamics, transient sensor readings, and collision avoidance among UAVs and aerial obstacles. By employing a virtual anchor point as a shared tracer and leveraging implicit formation control with only three UAVs, the proposed approach achieves robust, efficient, and accurate emitter localization.

The framework’s generalization capability was evaluated through quantitative studies, where training was performed using data from a single emitter and fixed wind conditions, and testing was conducted across various emitter locations and wind patterns. UAV behavior across different CPSL phases was further analyzed through detailed animations. Compared to the state-of-the-art fluxotaxis approach, the proposed MARL framework demonstrates superior performance in both localization accuracy and tracing efficiency.

This research opens several avenues for future work. One direction is to incorporate communication delays within the UAV network, requiring agents to use buffered sensor data from previous time steps. Another is to explore variable team sizes—enabled by decreasing costs of UAVs and sensors—which could improve robustness but also introduce new challenges such as increased computational complexity. Experimental validation is also part of our future plan, including the development of custom methane sensors, UAV-sensor integration, pilot training, and controlled-release field trials.

\section*{CRediT authorship contribution statement}
\textbf{Zhirun Li:} Data curation, Formal analysis, Methodology, Software, Validation, Visualization, Writing – original draft.
\textbf{Derek Hollenbeck:} Data curation, Formal analysis, Methodology, Software, Validation, Writing – original draft.
\textbf{Ruikun Wu:} Data curation, Formal analysis, Methodology, Software, Validation, Writing – original draft.
\textbf{Michelle Sherman:} Data curation, Formal analysis, Methodology, Software, Validation, Writing – original draft.
\textbf{Sihua Shao:} Conceptualization, Formal analysis, Funding acquisition, Methodology, Project administration, Supervision, Writing – original draft.
\textbf{Xiang Sun:} Conceptualization, Funding acquisition, Methodology, Project administration, Supervision, Writing – review and editing.
\textbf{Mostafa Hassanalian:} Conceptualization, Funding acquisition, Project administration, Supervision, Writing – review and editing.

\section*{Acknowledgment}
This work was supported by the National Science Foundation under Award under grant no. CNS-2323050 and CNS-2148178, where CNS-2148178 is supported in part by funds from federal agency and industry partners as specified in the Resilient \& Intelligent NextG Systems (RINGS) program.

\bibliographystyle{IEEEtran}
\bibliography{references}

\end{document}